\documentclass[journal]{IEEEtran}
\ifCLASSINFOpdf
\else
\fi
\usepackage{amsmath}
\usepackage{graphicx}
\usepackage{subfigure}
\usepackage{amsmath}
\usepackage[linesnumbered, ruled, vlined]{algorithm2e}
\usepackage{algpseudocode}
\usepackage{cite}
\usepackage{multicol}
\usepackage{multirow}
\usepackage{mathtools}
\usepackage{amssymb}
\usepackage{color}
\usepackage{bm}
\usepackage{indentfirst}
\usepackage{url}
\usepackage{amsthm}
\usepackage{nomencl}
\usepackage{booktabs}
\usepackage{cite}
\usepackage{siunitx}
\usepackage{marvosym}
\usepackage[numbers,sort&compress]{natbib}
\usepackage[font=small,skip=0pt]{caption}

\newtheorem{theorem}{Theorem}

\newtheorem{definition}{Definition}
\newtheorem{remark}{Remark}

\SetKwInOut{Input}{Input}
\SetKwInOut{Input}{Initialize}
\SetKwInOut{Output}{Output}

\begin{document}
\title{Optimal Sharing and Fair Cost Allocation of  Community Energy Storage}
	%On the Study of Community Energy Storage Models: Sizing, Valuation and Cost Allocation}
	%ADMM-based Optimal Sizing of Community Energy Storage Under Uncertainties}

\author{Yu~Yang,~\IEEEmembership{Student Member,~IEEE,}
	Guoqiang~Hu,~\IEEEmembership{Senior Member,~IEEE,}
	and~Costas~J.~Spanos,~\IEEEmembership{Fellow,~IEEE}% <-this % stops a space
	\thanks{This  work  was  supported  by  the  Republic  of  Singapore’s  National  Research  Foundation  through  a  grant  to  the  Berkeley  Education  Alliance  for  Research  in  Singapore
		(BEARS)  for  the  Singapore-Berkeley  Building  Efficiency  and  Sustainability  in  the
		Tropics  (SinBerBEST)  Program.  BEARS  has  been  established  by  the  University  of  California,  Berkeley  as  a  center  for  intellectual  excellence  in  research  and  education  in
		Singapore.}% <-this % stops a space
	\thanks{Yu Yang is with SinBerBEST, Berkeley Education 	Alliance for Research in Singapore, Singapore 138602 e-mail: (yu.yang@bears-berkeley.sg).}
	\thanks{Guoqiang Hu is with the School 	of Electrical and Electronic Engineering, Nanyang Technological University,
		Singapore, 639798 e-mail: (gqhu@ntu.edu.sg).}
	\thanks{Costas J. Spanos is with the Department of Electrical Engineering and 	Computer Sciences, University of California, Berkeley, CA, 94720 USA email: (spanos@berkeley.edu).}% <-this % stops a space
}

\maketitle

\begin{abstract}
This paper studies an energy storage (ES) sharing  model which is   cooperatively invested by multiple buildings  for harnessing  on-site renewable utilization and grid price arbitrage. 	
To maximize the economic benefits, we jointly consider the ES sizing, operation,  and cost allocation  via a coalition game formulation. 	
Particularly, we study a fair \emph{ex-post} cost allocation  based on \emph{nucleolus} which addresses fairness by minimizing the minimal  dissatisfaction of all the players. 
To overcome the exponential computation burden caused by the implicit characteristic function, we employ a constraint generation technique to gradually approach the unique \emph{nucleolus} by leveraging the sparse problem structure. 	
We demonstrate both  the fairness  and computational efficiency of the method through case studies, which are not provided by the existing  \emph{Shapley approach}  or  \emph{proportional method}. 
Particularly,  only a small fraction of characteristic function (less than 1\% for 20 buildings) is required to achieve the cost allocation versus the exponential information required by \emph{Shapley approach}. %Moreover, the proposed method can ensure fairness of the cost allocation whereas the \emph{proportional method} fails. 
Though there exists  a minor increase of computation over the \emph{proportional method}, the proposed method can ensure fairness while the latter fails in some cases. 
Further, we demonstrate  both the building-wise and community-wise economic benefits are enhanced with  the ES sharing model over the  individual ES  (IES) model. Accordingly, the  overall  \emph{value}{\footnote{The proportion of total electricity bill reduction relative to the ES capital cost.}} of ES  is considerably improved (about 1.83 times).
\end{abstract}


\IEEEpeerreviewmaketitle

\section{Introduction}

Energy storage (ES)  is  a  key  technology  for the world's transition to a sustainable, flexible and reliable energy system \cite{de2016value}.  
Based on the market applications, ES are commonly differentiated as grid-level and customer-level ES {\footnote{grid-level ES also refers to in-front of the meter (FTM) ES and customer-level ES refers to behind-the-meter ES.}}.  While grid-level ES are  connected to the distribution or transmission networks  by  system operators for ensuring energy supply quality,    customer-level  ES are usually deployed in consumer premise for harvesting economic benefits~\cite{Uitlity-scaleES}.  
Particularly, consumers paired with ES can cut their electricity bills by synchronizing  local volatile renewable supply with non-shiftable demand
and  responding  to grid  price variations~\cite{yu2020energy}.  
Over the years,   grid-level ES projects are being deployed globally \cite{Uitlity-scaleES},  whereas the  customer-level ES deployment is still largely impeded by the high ES capital cost and long payback period albeit the mandatory goals set by the governments~\cite{mallapragada2020long, hu2010optimal}. How to break or lower the barriers for customer-level ES deployment has become a critical  issue facing the energy system transitions.

%Mainly due to the  perception of  economic benefits,  there is a remarkable growth of behind-the-meter ES over the past few years. 
%However, though the economic benefits of customer-level ES are profound and the governments have set mandatory goals,   
% Currently, ES is still capital-intensive and economically questionable to the customers, which . 
%Therefore, how to break the barriers and accelerate the deployment of ES is a critical issue determining the pace of energy system transition. 
 
% to be addressed. Since energy storage is expensive and the operation of ES is greatly depends on customer variability, how to develop a viable business model is very important. 
% 
%One of the key enabling factor is to reduce the high upfront cost.

In recent years, \emph{sharing economy} has manifested in transportation and housing systems \cite{heinrichs2013sharing}. 
Naturally, such sharing-oriented solution is penetrating energy systems for bringing in new technologies,  such as ES  \cite{lombardi2017sharing, heinrichs2013sharing}. 
Particularly,  multiple users   can cooperatively  invest and share a central  ES~\cite{koirala2019community, zhu2019credit, koirala2018community}.
The added benefits of  sharing  ES over installing individual ES (IES) are comprehensive. 
\emph{First},  the utilization 
of ES can be increased by exploiting the complementary features of user production and consumption. 
For example, the users generally have  different  renewable generation  and load  patterns,  rendering  them to use the ES in a time division manner.
\emph{Second},  duplicate installation  cost can be avoided. The capital cost of ES deployment is composed of  battery cost and installation cost, the latter of which accounts for  20\%-50\%  \footnote{A Tesla Powerwall costs 7,600\$ before installation. However, accounting for the installation cost, a rough estimate of the Tesla Powerwall cost  \$9,600-\$15,600 for a full system installation. }. 
 \emph{Third},  additional benefit  from the economy of scale can be embraced.
 Similar to other goods or commodities,  customers are likely to  install  bulks of  ES at the wholesale price.  
 \emph{Last but not the least},  locating  a central ES  in  a community-shared space  can relieve  the space concern for the customers. 

The augmented benefits of  cooperative ES sharing are clear,  however how to reap them relies on a comprehensive business model which  is  a nontrivial combinatorial problem that integrates 
\begin{itemize}
\item ES sizing.  As  ES is capital-intensive, determining the appropriate  energy and power capacity is a prerequisite to maximize the economic benefits of the ES sharing.  

\item ES operation. The operation of the shared ES should coordinate the charging and discharging requirements  of all users. 

\item Cost allocation. Since  the  users are self-interested and independent stakeholders,  it is essential to study the fair cost allocation among the users. 
% involved. Reasonably, we refer to the fairness as cost or profit allocations that no customers have motivation to deviate or disrupt the cooperation by forming other sub-cooperations.  
\end{itemize}

In the literature,  a number of works  have studied  the business models for  cooperative ES sharing (see \cite{fleischhacker2018sharing, chakraborty2018sharing, lombardi2017sharing} and the references therein). However, most of them addressed  the ES sizing  (see \cite{bayram2015stochastic, kim2017optimal}, for examples) and  ES operation (see \cite{yao2015optimal, yao2017privacy}, for examples) separately. 
Nevertheless, they are interdependent and require to be coordinated so as to justify the high ES capital cost and maximize the economic benefits. Moreover, most existing works addressed cost allocation using rule-based cost allocation mechanisms which do not provide any notion of fairness  {\footnote{In this paper, we refer to fairness as satisfaction and we interchangeably use the word ``fairness" and ``satisfaction".  }}. 

\subsection{Main contributions}
\vspace{-1mm}
This paper studies a cooperative ES sharing model among multiple buildings, each of which  seeks economic benefits  from local renewable integration and  grid price arbitrage. Our main contributions are as follows.
\begin{itemize}
	\item We formulate the optimal ES sharing  integrating  optimal sizing,  operation and cost allocation as a coalition game. 
	%Particularly, our work can be regarded as an extension of \cite{chakraborty2018sharing} where we consider a more general market setting and account for renewable integration. 

	\item  We address the fair \emph{ex-post} cost allocation for ES sharing based on \emph{nucleolus}.
	In particular, we employ a constraint generation technique \cite{hallefjord1995computing} to overcome the exponential computation burden caused by the implicit characteristic function. 
	
	%is computationally efficient as it
	%only requires computing a small fraction of  coalition values, which is  in contrast to the general \emph{fair} cost allocation mechanism based on Shapley value that requires the exponential number of coalition values (characterized by a two-stage stochastic optimization problem). 
	
	\item  We demonstrate the superiority of the cost allocation over the existing   \emph{Shapley approach} and \emph{proportional method}  by providing both  fairness and  computational efficiency through case studies.  Particularly, only a small fraction of the characteristic function (less than 1\% for 20 buildings) is required to achieve the  cost allocation. 
	%Particularly,   we show the proposed method outperforms the existing  \emph{Shapley approach} and \emph{proportional method}  in fairness or computational efficiency, respectively. 
	
	\item We show the enhanced economic benefits of the ES sharing model  over the IES model both at the building-wise and community-wise.  Specifically, the ES sharing model yields  higher cost reduction to each committed building as well as the whole community.  Accordingly, the overall \emph{value} of ES is considerably  improved  (about 1.83 times).

	%by comparing with \emph{Shapley approach} and \emph{proportional method}  in a number of case studies. 
	%cost allocation mechanism as well as the economic benefits of  the CES  models.
	%
	%The numeric results imply that the proposed method can achieve a \emph{fair} cost allocation indicated by \emph{nucleous} with linear computation cost (w.r.t. the number of participants). 
	%
	%Moreover,  the numeric results demonstrate the enhanced  economic benefits  of  the CES model  both for individual buildings and the whole community microgrid (i.e., about $1.83$ times)  over  individual ES installation. 
\end{itemize}

The remainder of this paper is organized. 
In Section II, we review the related works. 
In Section III, we present the coalition game formulation.
In Section IV, we study the  \emph{fair} cost allocation.
In Section V, we study the fairness and computational efficiency of the cost allocation as well as  the economic benefits of the ES sharing model through  case studies.
In Section VI, we briefly conclude this paper.

\section{Related Works}
 In the literature,  a number of works have studied  ES sharing models from the perspective of sizing and operation.  

For ES sizing, \cite{bayram2015stochastic} proposed an analytical approach by using a Markovian fluid model  to capture the stochastic user demands where 
the ES was used as backup to ensure user demands under grid capacity limits.
% instead of storing renewable generation or achieving grid price arbitrage.
% The added economic benefits of the shared ES through coordinating  end-user behaviors  are not discussed.  
\cite{kim2017optimal}  studied the optimal sizing of shared  ES with the  objective to facilitate  photovoltaic (PV) utilization so as to minimize consumer cost.

The operation of  cooperative ES sharing generally addresses  the charging and discharging coordination among different users.  
Typically, \cite{zhu2019credit}  proposed a credit-based distributed algorithm  to manage  a central  ES  shared by a group of  cost-aware households.
The  discharging rate of the  shared ES is dynamically allocated among the households  
by their credits which characterize their accumulated stored energy in the shared ES.  
\cite{alskaif2017reputation} proposed an ES sharing model for  minimizing  the reputation-weighted energy cost of consumers, which are characterized by the proportions of  their renewable injection into the  shared ES over the historical time periods.   
Similarly, \cite{yao2015optimal,yao2017privacy} studied the control of a shared ES for optimizing the  total weighted cost  of a group of homes which reflects their  agreements on cost-saving priorities. 
\cite{zhang2020service} studied a shared  ES model working as an energy provider to serve consumers with elastic demand. The operation of  ES is managed by an aggregator obligated to minimize the total electricity bill for consumers. Particularly, a marginal service price model was deduced to charge the consumers. 
\cite{fleischhacker2018sharing} studied the optimal operation of a solar-plus-storage system shared across multiple consumers  for social welfare or profit maximization.  
%Generally, these works mainly focus on the ES operation and do not  address  the ES sizing issue. 
Generally, these works studied  the \emph{ad hoc} operation of a shared ES  for some specified objectives.  They  hardly addressed   the optimal ES sizing and capital cost allocation among the participants which are  important for practice.

 Besides, the sizing and operation of cooperative ES sharing  are mostly addressed separately in the literature,  however they should be jointly considered so as to justify the high ES capital cost and maximize the  economic benefits.
	%In a nutshell,  most  existing works  have separately addressed the  ES sizing (see \cite{bayram2015stochastic, kim2017optimal}) and ES operation (see \cite{yao2015optimal, yao2017privacy}) for ES sharing. 
	%However,  they are interdependent and  require  to be jointly considered so as  to maximize the economic benefits \cite{chakraborty2018sharing,liu126optimal}.  
	Moreover,  it is essential to study the fair cost allocation (i.e., ES capital cost) as the users  are self-interested and independent stakeholders.
	%discuss the allocation of ES capital cost or social benefits among the participants. 
	%Generally, the optimal ES sharing capitalizes on dynamical coordination of the users with complementary demands. However, this requires a paired cost allocation mechanism to \emph{fairly} distribute the co-created \emph{value} among the users.
	%We note that most of the  existing works have addressed  the  planning and operation  of  ES sharing separately. 
	%However, for capital-intensive ES resources, they should be jointly considered so as to  justify the high upfront cost or maximize the economic benefits.  Moreover, the optimal ES sharing necessitates the coordination of users with complementary demands. However, how to \emph{fairly} allocate the ES capital cost is a requisite.
	This paper works towards such objectives. 
Particularly, we study an  ES sharing model that is cooperatively invested and shared by multiple users to harness the economic benefits of grid price arbitrage as well as local renewable integration. 
	One close work  is  \cite{chakraborty2018sharing}  which studied the similar sharing paradigm  of ES,  and an analytical fair cost allocation formula based on  \emph{core} was  identified.  However, that work only considered grid price arbitrage for ES which is generally not enough to justify the high ES capital cost,  and the results (i.e., ES operation, sizing, and cost allocation) are restricted to the two-period market setting and cannot be extended to the case with time-dependent renewable generation.  Exceptionally, \cite{zhao2017pricing, zhao2019virtual} studied an ES sharing model that accounts for both price arbitrage and local renewable integration. 
	Whereas  those works focused on a different sharing paradigm in which a  third-party leads the ES sharing among its ``consumers" and pursues profit maximization.  Another related work  that shares the  similar structure of ours  by including the ES sizing, operation and cost allocation is  \cite{liu126optimal}. However, that work studied  the ES sharing across multiple electricity retailers. Besides,  the cost allocation  was rule-based  and did not address  the fairness due to the computational challenges.

\section{Coalition Game Formulation of ES Sharing}

\subsection{The Configuration of ES Sharing}
\begin{figure}
	\centering
	\includegraphics[width=3.0 in]{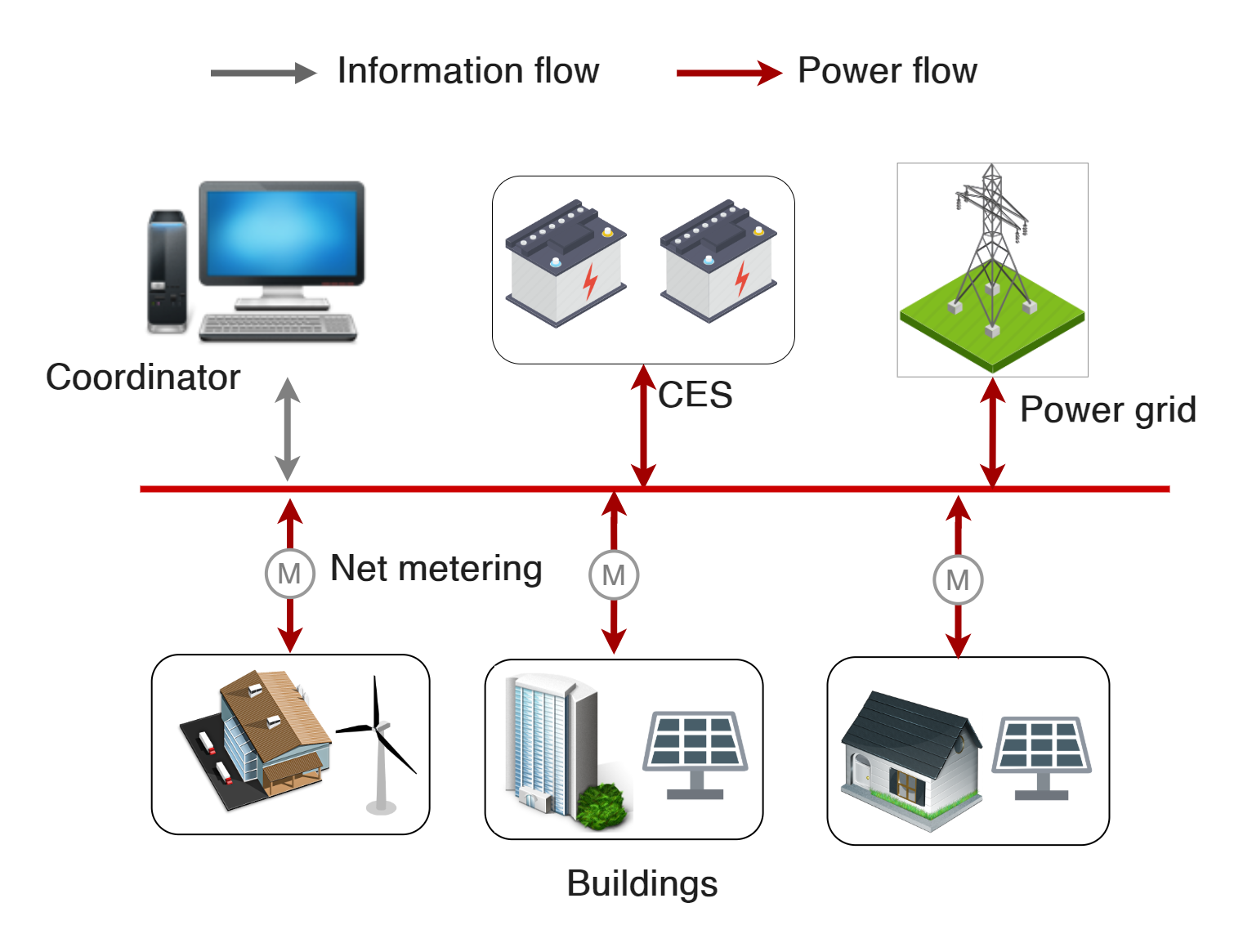}\\
	\caption{The configuration of a community energy storage (CES).}\label{system architecture}
\end{figure}

Fig. \ref{system architecture} shows  the configuration of the ES sharing model studied in this paper. 
%depicts the configuration of  a central ES shared among multiple buildings. 
We consider a grid-connected community composed of multiple buildings with on-site renewable generation (i.e., wind and solar power).  Wherein the buildings  cooperatively invest and share a community ES (CES) to harness   renewable utilization  and  grid  price arbitrage.  The buildings can charge the CES with  local renewable generation or the procured  electricity from the grid. 
Conversely, they can discharge it  to supply their demand  when required.  
Besides, we allow the buildings to sell  energy (i.e.,  renewable generation or discharged energy from the ES) back to  the grid. 
As the ES is shared by multiple buildings, we assume a central coordinator obligated to coordinate their charging and discharging behaviors of the buildings.  Supportively,  a net metering is attached to each building for monitoring the energy flow. Particularly, different buildings can charge and discharge simultaneously as the central coordinator only cares about the net power flow. In other word, if one building charges and another building discharges, there would exist some cancellations. However, each individual building can not charge or discharge at the same time due to the physical limits. 

%Since the renewable generation is volatile and it's usually not economic for the buildings to sell energy back to the grid, 
%the buildings are envisioned to store energy for future use.  Moreover, the buildings can gain some extra benefits by using ES to achieve price arbitrage: \emph{purchase electricity from the grid at lower price for later use with higher electricity price}.
%With such concerns, the buildings may solely invest their individual  ES (IES) or form as groups to cooperatively invest a common-owned  community ES (CES).    
%For IES,  the buildings will have exclusive access to the ES by undertaking the total capital cost of ES whereas for CES, it entails the buildings to cooperatively use and share the value of the ES resource.  
%Intuitively, for CES, all the building participants are inclined to use the ES maximally to reduce their electricity bills,  therefore to avoid negative competition,  there necessitates a central coordinator to dynamically distribute the ES resources across the participants so as to enhance the social benefits  (minimize the total energy cost) harnessed from the ES. 
% This is critical as only if  the total cost is minimized, the allocated cost for the buildings can be reduced.

We study the optimal ES sharing  model that encompasses the optimal  sizing, operation,  and \emph{ex-post} cost allocation.  
Considering  the computation burden of long-term planning (i.e., sizing),   we project the problem on a daily basis and study the problem in a discretized-time framework.

\subsection{Main Assumptions}
We clarify our main assumptions as below.  
\begin{itemize}
	\item[($A1$)] The purchase price of the grid is much higher than the  selling price.
	\item[($A2$)] The buildings do not share (or trade)  energy with each other through the shared ES.
	\item[$(A3)$] We only consider non-elastic demand of the buildings and demand response is not discussed. 
\end{itemize}

\begin{remark}
	($A1$) is a mild assumption  as  renewable generation is generally motivated to be consumed  locally instead of  fed into the grid as negative load. Meanwhile,  this setting can  prevent opportunistic folks   from disrupting  electricity market with ES: stocking  energy with lower price and selling back at a higher price.  
   ($A2$)  is imposed  to focus on the economic benefits of  ES sharing. However,  the model  can be extended to incorporate energy sharing. This supposes to enhance the economic benefits of  ES sharing further. 
 %($A3$) This paper doesn't consider demand response of the buildings. 
\end{remark}

%{\color{red}{degradation cost model}}
%\cite{farzaneh2019robust}

\subsection{Coalition Game Formation}
We formulate  the problem as a coalition game  \cite{saad2009coalitional}.
Coalition game is a branch of game theory that studies the cooperative behaviors of a group of rational agent, which accords with our settings.
Following the standard terminologies, we label the building players  by $\mathcal{N}: =\{1, 2, \cdots, N\}$. 
We study the problem over the discretized time slots  $\mathcal{T}: = \{0, 1, \cdots, T-1\}$.

%Since the ES capital cost is one-time charge and the buildings' demand generally reflects some daily cyclic features,  we project the long-term planning problem on a daily basis to reduce computation complexity.  We study the operation of the ES  in a discrete-time framework over the decision intervals $\mathcal{T}: = \{0, 1, \cdots, T-1\}$.

%A  coalition game is generally indicated by $(\mathcal{N}, \nu)$ where $\mathcal{N}$ represents the candidate players and $\nu$  denote the coalition value. 

\emph{ (i) Coalitions}:  For a community composed of $N$ buildings,  we refer to any subset of   buildings  $ \mathcal{S} \subseteq \mathcal{N}$ cooperatively sharing  an ES as an ES coalition or sub-coalition $\mathcal{S}$. 
% Therefore, the buildings in the community microgrid may form a variety of coalitions indicated by $\mathcal{S}_1, \mathcal{S}_2, \cdots, \mathcal{S}_l$, and we have $\mathcal{S}_i \cap \mathcal{S}_j = \emptyset$, $\forall i\neq j$ and 
%$\cup_{i = 1} ^l \mathcal{S}_i = \mathcal{N}$. 

\emph{(ii) Coalition value and characteristic function}: Coalition value $\nu(\mathcal{S})$ quantifies the economic benefit of an ES coalition $\mathcal{S}$. 
We indicate a $N$-player  coalition game by  $(\mathcal{N}, \nu)$, where function $\nu: 2^N \rightarrow \mathbb{R}$ represents the characteristic function of coalition game that assigns \emph{value} $\nu(\mathcal{S})$  to each sub-coalition $\mathcal{S} \subseteq \mathcal{N}$.  Particularly,  the number of sub-coalitions  $ \mathcal{S} \subseteq \mathcal{N}$ grows exponentially with the scale, i.e., $O(2^N)$. 

In this paper, we characterize the \emph{value} of an  ES  coalition $\nu(\mathcal{S})$ by  the total cost: the total electricity bills  plus  the ES capital cost. 
When we use a collection of representative scenarios to capture the  patterns of renewable generation and building demand,  the coalition  value $\nu(\mathcal{S})$ can be characterized by  a two-stage stochastic optimization problem that couples the optimal ES sizing  and  operation: 

\vspace{-3mm}
{\small 
	\begin{subequations} \label{main_problem}
		\begin{alignat}{4}
		%&(\mathcal{P}):\notag \\
		(\bm{\mathcal{P}}):~&\label{UperLevel}\nu(\mathcal{S}) = \min_{E_{\mathcal{S}}, P_{\mathcal{S}} \geq 0}  c(\bm{x}_{\mathcal{S}})+\sum_{\omega \in \Omega} \rho_\omega g(\bm{x}_{\mathcal{S}}, \zeta_\omega)  \tag{$\mathcal{P}_U$}\\
		\label{eq:LowerLevel}\quad &g(\bm{x}_{\mathcal{S}},  \zeta_\omega)=\min_{\bm{y}^{\omega}_i,  i \in \mathcal{S} } g(\bm{x}_{\mathcal{S}},  \bm{y}^{\omega}, \zeta_ \omega) \tag{$\mathcal{P}_L$}\\
		%&\textrm{subject to}  \notag\\
		\textrm{subject to:} &\label{eq:1a} ~~ \bm{y}^\omega_i  \in \mathcal{Y}^\omega_i, ~ i \in \mathcal{S}. \\
		&\label{eq:1b} ~~  \sum_{i \in \mathcal{S}} e^{b, \omega}_{i, t} \leq E_{\mathcal{S}},  t \in \mathcal{T}. \\ 
		&\label{eq:1c}~~ \sum_{i \in \mathcal{S}} p^{ \text{ch}, \omega}_{i, t} \leq P_{\mathcal{S}},  t \in \mathcal{T}. \\
		&\label{eq:1d}~~ \sum_{i \in \mathcal{S}} p^{ \text{dis}, \omega}_{i, t} \leq P_{\mathcal{S}},  t \in \mathcal{T}.
		\end{alignat}
\end{subequations} }
where $i$ and $t$ are building and time indices.  $\bm{x}_{\mathcal{S}} = (E_{\mathcal{S}}, P_{\mathcal{S}})$ denotes the ES capacity: energy capacity  $E_{\mathcal{S}}$ (in \si{\kilo\watt\hour}) and power capacity  $P_{\mathcal{S}}$ (in \si{\kilo\watt}).   $\omega$, $\zeta_\omega$, $\Omega$  and $p_{\omega}$ represent scenario indices, scenario realizations, scenario collection and scenario probabilities. $\bm{y}^{\omega}_i$ denotes the operating strategy for building $i$ under scenarios $\omega$,  which includes  the charging/discharging of the ES and energy trading with the grid. Accordingly,  $\mathcal{Y}^{\omega}_i$ indicates the set of admissible strategies. 

\begin{itemize}
	
	\item The first-stage objective \eqref{UperLevel} captures  the total cost which is equal to the ES capital cost $c(\bm{x}_{\cal S})$ plus the weighted  operation cost  $\sum_{\omega \in \Omega} p_\omega g(\bm{x}_{\mathcal{S}}, \zeta_\omega)$.  For the ES capital cost, we capitalize on an amortized price model \cite{harsha2014optimal, zhao2019virtual}:  $c(\bm{x}_{\mathcal{S}})=k_p P_{\mathcal{S}} + k_e  E_{\mathcal{S}}$.  $k_p$  $(s\$/\si{\kilo\watt\hour})$ and $k_e$ $(s\$/\si{\kilo\watt})$  are the amortized ES capacity price which are obtained according to the projected 
		ES price 
		$100$\EUR\si{\per{\kilo\watt\hour}} and $300$\EUR\si{\per{\kilo\watt}} by 2025 \cite{pandvzic2018optimal}.  
	%The calculation of the amortized ES price refers to \textbf{Appendix} A. 

	\item The second-stage objective \eqref{eq:LowerLevel} characterizes the optimal operation cost $g(\bm{x}_{\cal S},  \bm{y}^{\omega}, \zeta_ \omega)$ for each scenario $\omega$,  which is subject to the ES capacity $\bm{x}_{\cal S}$,  the scenario realization $\zeta_\omega$,  and the building operating strategies $\bm{y}^\omega=[y^\omega_i], \forall i \in \mathcal{S}$.  
	%This paper studies the energy market characterized by the contracted power capacity tariff and Time-of-Use (TOU) tariff. 
	The operation cost  consists of  electricity purchase cost and  the revenue of selling energy to the grid. We use  $c^{\text{g+}}_t$, $c^{\text{g-}}_t$  and $c^{g, \max}$  to indicate the purchase, selling and demand charge price. 
	$p^{\text{g+}, \omega}_{i, t}$ and $p^{\text{g-}, \omega}_{i, t}$ denote the procured and sold  energy of building $i$ over period $t$.
	$p^{g, \max, \omega}_i$ characterizes the peak demand over the billing cycle. 
	We  have the operation cost:
	
	\vspace{-3mm}
	{\small 
		\begin{equation} \label{operation cost}
		\begin{split}
		g(\bm{x},  \bm{y}^{\omega}, \zeta_ \omega)=& \sum_{i \in \mathcal{S}} \Big\{ \sum_{t \in \mathcal{T} }  \big( c^{\text{g+}}_t p^{\text{g+}, \omega}_{i, t}-c^{\text{g-}}_t  p^{\text{g-}, \omega}_{i, t}\big) \\
		&+c^{g, \max} p^{g, \max, \omega}_i \Big\}  
		\end{split}
		\end{equation} }

	\item We define the operation strategy of building $i$  as $\bm{y}^{\omega}_i=[p^{\text{ch}}_{i, t}, p^{\text{dis}}_{i, t}, e^{\text{b},  \omega}_{i, t},  p^{\text{g+}, \omega}_{i, t},  p^{\text{g-}, \omega}_{i, t}, $ $p_i^{g, \max, \omega}]$ which includes   the charging and discharging energy: $p^{\text{ch}}_{i, t}, p^{\text{dis}}_{i, t}$,   the stored energy:  $e^{\text{b},  \omega}_{i, t}$,  the procured and sold energy from/to the grid:  $p^{\text{g+}, \omega}_{i, t},  p^{\text{g-}, \omega}_{i, t}$ and the peak demand over the billing cycle: $p_i^{g, \max, \omega}$. 
	Correspondingly,  the set of admissible operation strategies   $\mathcal{Y}^{\omega}_i$  is  constituted by
	
	\vspace{-3mm}
	{\small  
		\begin{subequations}
			\begin{align}
			&\label{eq:4a} 0 \leq p^{\text{ch}}_{i, t}  \leq p^{\text{ch}, \max},  \\
			&\label{eq:4b} 0 \leq p^{\text{dis}}_{i, t}  \leq p^{\text{dis}, \max}, \\
			%&\label{eq:4c} p^{\omega}_{i, t}=p^{\text{ch}, \omega}_{i, t} -p^{\text{dis}, \omega}_{i, t},  \\
			& \label{eq:4c} e^{\text{b}, \omega}_{i, t+1}=e^{\text{b}, \omega}_{i, t}+p^{\text{ch}, \omega}_{i, t} \eta^{\text{ch}}- p^{\text{dis}}_{i, t} / \eta^{\text{dis}}, \\
			& \label{eq:4d} e^{\text{b}, \omega}_{i, t} \geq  0,  \\
			& \label{eq:4e} p^{ \text{g+}, \omega}_{i, t}- p^{\text{g-}, \omega}_{i, t} = p^{\text{ch}, \omega}_{i, t} -p^{\text{dis}, \omega}_{i, t}+p^{\text{d}, \omega}_{i, t}-p^{\text{r}, \omega}_{i, t}, \\
			&\label{eq:4f} p^{\text{g+}, \omega}_{i, t} , p^{\text{g-}, \omega}_{i, t}\leq p_i^{g, \max, \omega},  \\
			& \label{eq:4g} p^{\text{ch}, \omega}_{i, t}  p^{\text{dis}, \omega}_{i, t} = 0,\\
			& \label{eq:4h} p^{\text{g+}, \omega}_{i, t} p^{\text{g-}, \omega}_{i, t} = 0,~\forall t \in \mathcal{T}. 
			\end{align}
	\end{subequations}}
	where   constraints \eqref{eq:4a}-\eqref{eq:4b} model the charging and discharging rate limits. 
	%$p^{\text{ch}, \omega}_{i, t}$  and $p^{\text{dis}, \omega}_{i, t}$  indicate the charging and discharging power of building $i$.
	%Accordingly, 
	%The constants $p^{\text{ch}, \max}$ and  $p^{\text{dis}, \max}$ denote the maximum charging and discharging power. 
	Constraint \eqref{eq:4c} tracks the stored energy for each building subject to  the charging and discharging efficiency $\eta^{\text{ch}}, \eta^{\text{dis}}$. 
	In this formulation, we do not consider energy sharing,  thus each building can not over deplete its stored energy  as imposed by constraint \eqref{eq:4d}. Constraint  \eqref{eq:4e}  models the instantaneous  balance of supply and demand of  each building,  with
	% $p^{\text{g+}, \omega}_{i, t}$ and $p^{\text{g-}, \omega}_{i, t}$ denote the procured and sold  power with the grid. 
	$p^{\text{d}, \omega}_{i,t}$ and $p^{\text{r}, \omega}_{i,t}$ denoting  the non-elastic demand and local renewable generation. %The energy trading (procure or sell) of each building 
	Constraint \eqref{eq:4f} captures peak demand over the billing cycle. 
	Particularly, the complementary constraints \eqref{eq:4g}-\eqref{eq:4h} enforce the physical limits of non-simultaneous  charging (purchasing) and discharging (selling). 
\end{itemize}

%are complementary constraints enforcing each building not charge/discharge or  purchase/sell energy at the same time.  

%Each building participant can charge and discharge the CES by their requirements as long as their stored energy is not depleted indicated by constraint. 

%$e^{\text{b}, \omega}_{i,t}$  tracks  the  stored energy of building $i$  in the CES . 
%$\eta^{\text{ch}}$ and $\eta^{\text{dis}}$ represent the ES roundtrip efficiency.
%$p^{\text{g+}, \omega}_{i, t}$ and $p^{\text{g-}, \omega}_{i, t}$ denote the procured and sold  power to the grid. 
%$p_i^{g, \max, \omega}$ indicates the peak demand of building $i$ which corresponds to the demand charge paid to the grid. 
%$p^{\text{d}, \omega}_{i,t}$ and $p^{\text{r}, \omega}_{i,t}$ denote the non-elastic demand and local renewable generation of building $i$. 
%Constraints \eqref{eq:4a}-\eqref{eq:4b} impose the physical limits on the charging and discharging power of building $i$. 

%maximum charging and discharging power limits $p^{\text{ch}, \max}$ and 
%$p^{\text{dis}, \max}$  on  building $i$.  
%Constraint \eqref{eq:4f} ensures the buildings trade (purchase or sell) energy with the grid within the contracted power capacity.  Constraints \eqref{eq:4h}-\eqref{eq:4i} are complementary constraints enforcing each building not charge/discharge or  purchase/sell energy at the same time.  

Note that problem $(\mathcal{P})$ is   non-linear and non-convex   due to the  presence of  complementary constraints \eqref{eq:4g}-\eqref{eq:4h},  making it computationally intractable with \emph{off-the-shelf} solvers.
However, considering the  ES efficiency, i.e., $\eta^{\text{ch}}, \eta^{\text{dis}} <1$,  and the  grid price setting, i.e., $c^{\text{g+}}_t > c^{\text{g-}}_t$, the complementary constraints can be relaxed without affecting the optimal solution.  We refer the readers to an illustrative proof in \textbf{Appendix} A. 
With constraints \eqref{eq:4g}-\eqref{eq:4h} relaxed,  there only exist  linear constraints.
Besides,  we note that the two-stage problem has a  min-min structure,  making  it possible to be  converted to the  single-stage convex problem  \eqref{eq:MainProblem}  that  can be tacked 
	%However, the two-stage structure  poses challenges to apply
	by some existing commercial solvers (e.g., CPLEX). 
%To address this issue,  we convert the two-stage problem to a single-stage one:
\begin{equation} \label{eq:MainProblem}
\begin{split}
(\bm{\mathcal{P}^{'}}):& \nu(\mathcal{S}) = \min   c(\bm{x}_{\mathcal{S}})+\sum_{\omega \in \Omega} \rho_\omega  g(\bm{x},  \bm{y}^{\omega}, \zeta_ \omega) \\
\text{subject~to:}~~&  \eqref{eq:1b}-\eqref{eq:1d}. ~\eqref{eq:4a}-\eqref{eq:4g},~\forall i \in \mathcal{S}. \\
& \text{var:}~ E_{\cal S},  P_{\cal S}, \bm{y}^{\omega}_i, \forall i \in \mathcal{S}. 
\end{split}
\end{equation}

%According, the objective function in the lower level problem \eqref{eq:LowerLevel} can be reformulated as 
%\begin{equation}
%\begin{split}
%g(\bm{x},  \bm{y}^{\omega}, \zeta_ \omega)=& \sum_{i \in \mathcal{N}} \Big\{ \sum_{t \in \mathcal{T} }  \big( c^{\text{g+}}_t p^{\text{g+}, \omega}_{i, t}-c^{\text{g-}}_t  p^{\text{g-}, \omega}_{i, t}  \big) \\
%&+c^{g, \max} p^{g, \max, \omega}_i \Big\} 
%\end{split}
%\end{equation}
%where the last part denotes the degradation cost,, we refer the readers to \cite{farzaneh2019robust} for details of the model. 
%In this subsection, we study the characteristics of the CES coalition game. 
%We have the following main results. 
\vspace{-4mm}
\subsection{ES Coalition Game Properties}
In this part, we study the characteristics of the ES coalition game. We have the following main result.

\begin{theorem} \label{theorem1}
	The ES  coalition game $(\mathcal{N}, \nu)$ is subadditive, i.e., $\nu(\mathcal{S}^1 \cup \mathcal{S}^2) \leq \nu(\mathcal{S}^1 ) + \nu(\mathcal{S}^2), ~\forall \mathcal{S}^1, \mathcal{S}^2\subseteq \mathcal{N}, \mathcal{S}^1 \cap \mathcal{S}^2 = \emptyset. $
	%	\begin{equation*}
	%	\begin{split}
	%	\nu(\mathcal{S}^1 \cup \mathcal{S}^2) \leq \nu(\mathcal{S}^1 ) + \nu(\mathcal{S}^2), ~\forall \mathcal{S}^1, \mathcal{S}^2 \subseteq \mathcal{N}, \mathcal{S}^1 \cap \mathcal{S}^2 = \emptyset. 
	%	\end{split}
	%	\end{equation*}
	%	where $\mathcal{S}_1$ and $\mathcal{S}_2$ indicates CES coalitions 
	
	% Therefore, the buildings in the community microgrid may form a variety of coalitions indicated by $\mathcal{S}_1, \mathcal{S}_2, \cdots, \mathcal{S}_l$, and we have $\mathcal{S}_i \cap \mathcal{S}_j = \emptyset$, $\forall i\neq j$ and 
	%$\cup_{i = 1} ^l \mathcal{S}_i = \mathcal{N}$. 
\end{theorem}

\begin{remark}
	We defer the proof  to  \textbf{Appendix} B. \textbf{Theorem} \ref{theorem1} implies that it won't be worse off for two  disjoint groups of buildings to merge and share a single ES.
	Fundamentally, the overall economic benefits  can be enhanced through merging. 
	Therefore the buildings within a community are inclined to form a grand  ES coalition $\mathcal{N}$ to maximize economic benefits.  
	%Therefore, this paper focuses on the grand coalition $(\mathcal{N}, \nu)$. 
	
	%Particularly,  we deduce that it will be economical beneficial for the buildings to cooperatively share a central CES compared with installing their own IES.
\end{remark}

\vspace{-5mm}
\section{Cost Allocation based on Nucleolus}

In Section II-D, we have proved the enhanced overll benefits  of the ES sharing model over IES model. 
%Specifically, the overall cost of the building participants can be reduced through sharing ES  over  individual ES installation. 
However, the building-wise gains  relies on the \emph{ex-post} cost allocation, which is supposed  to be fair  to ensure stable cooperation:  all participants are satisfied  and have no motivations  to deviate or disrupt the cooperation. 
%A  cost allocation that satisfies all building participants is a requisite  to stabilize the ES coalition,   otherwise some   may deviate or disrupt the cooperation.

% A \emph{fair} cost allocation mechanism requires to allocate the cost (or payoff) among the building participants  based on their \emph{marginal contribution}. Generally,  the  \emph{marginal contribution} of building $i$ to a coalition $\mathcal{S}$ can be quantified by the increased coalition value by excluding  building $i$ from the coalition, i.e.,  $\nu( \mathcal{S}\setminus\{i\})- \nu(\mathcal{S})$. 

%The cost allocation regarding the CES coalition game is a non-trivial task due to the heterogeneous  building participants. 
% To address this issue, this paper develops a cost allocation mechanism based on  the concept of  \emph{nucleolus} for coalition game. Specifically, to overcome the computational challenges,  a constraint generation approach is adopted to reduce the computation burden of characteristic functions for coalitions. 
%
% 
% \subsection{Preliminaries}

There exist multiple 
% various principles regarding the cost allocation of coalition (cooperative) games. For example, one may want to distribute the cost in a way that is ``fair" or ``stable". These 
%principles are called 
\emph{solution concepts}  regarding  fair cost allocation of  coalition game \cite{schulz2013approximating}. One prominent one  is   \emph{core}  \cite{gillies1959solutions, saad2009coalitional}. 
%which  characterizes  \emph{stable} cost allocations, i.e.,  no  groups of players has incentives to deviate from the current coalition 
%There are some standard procedures to compute the \emph{core}. 
Normally,  computing  \emph{core} corresponds to  a NP-complete linear programming (LP) that depends on the  entire characteristic function.  For example, for a $N$-player coalition game,  it generally requires to solve a LP problem with ($2^N-1$) constraints  associating with $O(2^N-1)$ coalition \emph{value}. 
This is  a non-trivial task due to the exponential computation burden. 
%As the number of sub-coalitions grow exponentially with the scale (i.e., $O(2^N)$), it  takes exponential time to compute the entire characteristic function. 
Besides,   the existence and uniqueness of   \emph{core}  is another general concern for practice  \cite{saad2009coalitional}.
%some other  flaws in practice \cite{saad2009coalitional}: \emph{i)} the \emph{core} of a coalition game  can be empty;  and \emph{ii)} the \emph{core} may not be unique.%;  and \emph{iii)} the core can be unfair to some players. 

Another primary \emph{solution concept} is  \emph{Shapley} value with the main  idea  of  distributing  the co-created  \emph{value}  by  the players' marginal contributions. 
%Specifically,  the marginal contribution of  player  $i$ to an existing coalition $\mathcal{S}$ can be quantified by the increased coalition \emph{value}:   $\nu( \mathcal{S}\!\setminus\!\{i\})\!-\!\nu(\mathcal{S})$. 
One  main advantage of  \emph{Shapley}  over \emph{core}  is the existence and uniqueness.  
However, it also suffers computation intensity from  the entire characteristic function. %marginal contributions  of  players.
%, which takes exponential time. 

%In general, there is no direct relationship between the \emph{core} and \emph{Shapley value}. However, for some special cases such as convex games, the Shapley value lies in the \emph{core} \cite{saad2009coalitional}. 
%Intuitively, the main computation burden regarding Shapley value based cost allocation approach is to characterize the marginal contribution of each individual players.   Similar to other applications of coalition games, this is a non-trivial and computationally intensive task for the CES coalition game.  

\emph{Nucleolus} is  another essential \emph{solution concept} which  pursues fairness  by 
minimizing the dissatisfaction of all players \cite{schmeidler1969nucleolus}.  %maximal
Reasonably, a  cost allocation  can be viewed as fair  if all players are satisfied (i.e., the maximum dissatisfaction of all players is non-positive). 
%The motivation behind  \emph{nucleous} is to  search for a cost allocation mechanism that satisfy all players.  
Besides, the nonempty and uniqueness  property of \emph{nucleolus}  is preferable, motivating  us to study the  cost allocation of  ES  sharing based on \emph{nucleolus}. Nevertheless,  the computation challenge is yet to be addressed as computing \emph{nucleolus} is as or even more difficult than the other \emph{solution concepts}.

\subsection{Definitions} 

We first  introduce  the main  definitions. % associated with cost allocation, \emph{core} and \emph{nucleolus}. % and \emph{core}.

\begin{definition}
	(Cost allocation) We use  vector $\bm{x} \! \in\!\mathcal{R}^{N}$ to denote a  cost allocation of coalition game $(\mathcal{N}, \nu)$, where  the entry $x_i$  denotes  the allocation to player $i$.
\end{definition}

\begin{definition}
	(Imputation) An imputation $\bm{x} \in \mathbb{R}^N$ is a cost allocation for a  grand coalition  $(\mathcal{N}, \nu)$ which is both efficient
	and individually rational. We have the set of imputations:  
	\[ \mathcal{I} = \Big\{  \bm{x} \in \mathbb{R}^N: x(\mathcal{N}) = \nu(\mathcal{N}) ~\textrm{and}~ \nu(\{i\}) \geq x_i, \forall i \in  \mathcal{N} \Big\} \]
	where we define $x(\mathcal{S})= \sum_{i \in \mathcal{S}} x_i, \forall \mathcal{S} \subseteq \mathcal{N}$.
\end{definition}

\begin{definition}\label{def:core}
	(Core) Core  refers to  the  imputation that no subsets of players has  incentives to deviate from the grand coalition $(\mathcal{N}, \nu)$. The set of core is defined as
	\begin{equation*} 
	\mathcal{C}= \Big\{  \bm{x} \in \mathcal{I}:~ x(\mathcal{S})  \leq \nu(\mathcal{S}), \forall \mathcal{S} \subseteq \mathcal{N} \Big\}
	\end{equation*}
\end{definition}

%From the \textbf{Definition} \ref{def:core}, we note that the cores are some cost allocations that is \emph{efficient}, i.e., $\sum_{i\in\mathcal{N}} x_i = \nu(\mathcal{N})$ and \emph{rational}, i.e., no subset of players would be better off by deviating from the grand coalition.  Clearly, whenever one is able to find a cost allocation that lies in the core, then the grand coalition is stable and optimal for the coalition game. However, the cores of coalition games are not always guaranteed to exist and the core may be empty. 
% Besides, even the cores do exist,  the search for the cores is generally NP-compete \cite{conitzer2003complexity} as it corresponds to a  LP  problems  that involves an enormous of  linear constraints. 

%Beside, we  display some important definitions related to  \emph{nucleolus}. 
%The specific definitions are displayed below. 
% Another important solution concept of cost allocation is \emph{nucleolus} which can compensate the feasibility of core issue. Specifically, the basic motivation behind \emph{nucleolus} is to find a cost allocation that minimize the dissatisfactions of the players on the coalitions.
% Before we give the specific definition of \emph{nucleolus}, we first give the definition of \emph{lexographical order} and \emph{the excess of coalition}. 
\begin{definition}
	(Lexographical order)  Assume two $N$-dimensional vectors $\bm{y}, \bm{z} \in \mathbb{R}^N$ with elements arranged in non-increasing order, i.e., $y_i \geq y_j$  and $z_i \geq z_j$ if $i <j$. We claim  vector $\bm{y}$ is 
	lexographically smaller than  vector $\bm{z}$, i.e., $\bm{y} \prec_{\text{lex}} \!\! \bm{z}$,  if $\exists k \!< \!N $ that $y_i = z_i, \forall i < k$  and $y_k\! <\! z_k$ (if $k=1$, we have $y_1 < z_1$). 
\end{definition}

\begin{definition}
	(Excess of coalition)  For a  given cost allocation $\bm{x} \in \mathbb{R}^{|\mathcal{S}|}$, the excess of coalition $\mathcal{S}$  is defined as 
	\[ e(\bm{x}, \mathcal{S}) =x(\mathcal{S}) - \nu(\mathcal{S})\] 
\end{definition}

For a coalition game characterized by cost minimization, the \emph{excess of coalition} $e(\bm{x}, \mathcal{S})$ can be interpreted as the \emph{dissatisfaction} of  coalition $\mathcal{S}$ with the cost allocation $\bm{x}$.  Since there exists a group of players,  the \emph{dissatisfaction} of players regarding a specified cost allocation $\bm{x} \in \mathbb{R}^N$ correspond to a \emph{excess of coalition}  vector. 
To address this issue,  \emph{nucleolus}  is defined by  the  lexographical order of \emph{excess of coalition} vectors.   
More specifically, \emph{nucleolus} is the \emph{imputation}  with the lexicographically minimal \emph{excess of coalition} vector. The formal definition is given below.

\begin{definition} \label{def:nucleolus}
	(Nucleolus)   %The nucleolus of a coalition game is the lexicographically minimal  imputation $ \bm{x} \in \mathcal{I}$. Specifically,  f
	For a coalition game $(\mathcal{N}, \nu)$,   let  $O(\bm{x}) \in \mathbb{R}^{2^{N}-1}$ be the \emph{excess of the coalition} vector for  cost allocation $\bm{x} \in \mathbb{R}^N$ (imputation) arranged in a non-increasing order, i.e., $O_i(\bm{x}) \geq O_j(\bm{x}), \forall i < j$, then a cost allocation $\bm{x} \in \mathbb{R}^N$ is the nucleolus if  we have  
	\[ O(\bm{x} ) \prec_{\text{lex}}O(\bm{x}^{'}), \forall \bm{x}^{'} \in  \mathcal{I} \setminus \{\bm{x}\}. \]
\end{definition}

\vspace{-15pt}

\subsection{An algorithm to find the nucleolus}
%We  first refer the readers to our extended version \cite{yang2020optimal} for 
%The specific definitions of  \emph{cost allocation},  \emph{imputation}, \emph{core}, \emph{lexographical order}, \emph{excess of coalition} and \emph{nucleolus} mentioned below can refer to our extended version 
%\cite{yang2020optimal}. 

From the definition,  we note that \emph{nucleolus}  is  always nonempty and  unique. 
	Particularly,   \emph{nucleolus} will locate  in the  \emph{core} if the latter is nonempty \cite{hallefjord1995computing}. 
However,  computing \emph{nucleolus} is nontrivial as it corresponds to searching for the lexicographically minimal  \emph{excess of coalition} vector.  More specifically, we are required to solve a sequence of lexicographically minimization problems, i.e.,  first   identifying  the set of cost allocations $\bm{X}_1$ that minimize $O_1(\bm{x})$,  and then   minimize $O_2(\bm{x})$ over $\bm{X}_1$,  where  $O_i(\bm{x})$ denotes the $i$-th entry of the \emph{excess of coalition} vector $O(\bm{x})$ (see  \textbf{Definition} \ref{def:nucleolus}). These episodes are carried forward until the unique \emph{nucleolus} is approached. 
%As the \emph{nucleolus} always exist and is unique, this paper studies a  \emph{fair} \emph{ex-post} cost allocation for the CES coalition game based on \emph{nucleolus}.  
Particularly,  we note that the entire  characteristic function is required in each episode to compute the \emph{excess of coalition} vector. 
This is computationally intensive for ES coalition game as the characteristic function is  implicit and
characterized by stochastic optimization problems. 
For example, for an ES coalition with $20$ buildings, we are required  to solve the stochastic optimization problem \eqref{eq:MainProblem}   $2^{20}-1$ (more than $10^7$) times to identify the  characteristic function.

To overcome the computational challenges, this section develops an algorithm to search for \emph{nucleolus}  of  the  ES coalition game by employing  a constraint generation technique \cite{hallefjord1995computing}. The main idea is  spurred by the
uniqueness of \emph{nucleolus} and the underlying sparse structure of  problem. 
Specifically for a $N$-player coalition game,  the cost allocation  corresponds to solving  a  sequence of  LPs (i.e., lexicographically minimization)  with $N$ decision variables  subject to $2^N-1$ linear constraints. 
This implies at most  $N$ of  $2^{N}-1$  constraints are \emph{binding} at the optima.
Intuitively, if the  $N$ \emph{binding} constraints are known \emph{a priori}, only $N$ coalition \emph{value} are actually required. 
Indeed, the essential idea of the proposed cost allocation  is to identify such critical  constraints using constraint generation technique. % to approach  the unique \emph{nucleolus}.
%This can greatly decrease the computation burden. 
%Unfortunately, there is no way to find out such constraints in advance.  However, the fundamental idea is still informative. 

%This section develops an \emph{effective} and \emph{computationally efficient} cost allocation mechanism to find the \emph{nucleolus} based on constraint generation technique \cite{hallefjord1995computing}. 
To be noted,  the algorithm to search for \emph{nucleolus}  is constituted by multiple episodes. 
We first starts with computing  the \emph{least core} \cite{schulz2013approximating}.  After that a sequence of   lexiographically minimization  problems are solved. In each episode, we capitalize on a constraint generation technique to identify the dissatisfied coalitions with the current cost allocation. 
The algorithm will terminate until  we encounter the unique  \emph{nucleolus}.   In the subsequent, we introduce the details. 
%This method first starts with the \emph{least core} \cite{schulz2013approximating}.
% and then alternatively performing the \emph{lexicographically optimization} and \emph{constraint generation} procedure. 

\subsubsection{Computing  least core}
The \emph{least core} of coalition game $(\mathcal{N}, \nu)$  is the solutions of  the LP problem \cite{schulz2013approximating}:
{\small 
	\begin{equation}\label{eq:least core}
	\begin{split}
	z^{*}=\min z & \\
	\text{subject~to:}~ &  x(\mathcal{N}) = \nu(\mathcal{N}) \\
	& x(\mathcal{S})-\nu(\mathcal{S}) \leq z, ~\forall \mathcal{S} \subseteq\mathcal{N}\!\setminus\! \{\emptyset, \mathcal{N} \}. 
	\end{split}
	\end{equation}
}

Problem \eqref{eq:least core}  is equivalent to {\small $ z^{*} = \min\limits_{\bm{x} \in \mathcal{I}} \max\limits_{\mathcal{S} \subseteq \mathcal{N} \setminus \{\emptyset, \mathcal{N} \} } e(\bm{x}, \mathcal{S})$.} We note that \emph{least core}   is  the  set of \emph{imputations} with the minimum maximum \emph{excess of coalition}, i.e., 
$\min_{\bm{x} \in \mathcal{I}} O_1(\bm{x})$.  Therefore, the \emph{least core}  includes  \emph{nucleolus}  based on the definition.

%but  the converse is not necessary. 
%Besides, we note that  both the \emph{least core} and the \emph{nucleolus} are closely related to the \emph{excess vector of coalitions} $O(\bm{x})$. 
%Moreover, according to \textbf{Definition} \ref{def:nucleolus}, we know 

 \emph{An example}: assume a $2$-player coalition game with the \emph{excess of coalition} vectors
	$O(\bm{x})=[5, 3, 3], O(\bm{y}) = [ 5, 4, 2], $ and $O(\bm{z}) = [ 6, 4, 1]$ for the cost allocation $\bm{x}, \bm{y}, \bm{z}$.  We can figure out  the  \emph{least core}  $\bm{x},  \bm{y}$  and  the  unique \emph{nucleolus}  $\bm{x}$ because we have $O_1(\bm{x}) = O_1(\bm{y}) \leq O_1(\bm{z})$ and $O_2(\bm{x}) < O_2(\bm{y})$.

Clearly,  \emph{least core}  is well-defined and can be used to narrow the search scope of \emph{nucleolus}. 
Nevertheless,   computing  \emph{least core} by solving problem \eqref{eq:least core} is computationally intensive as it requires the entire characteristic function: $\nu: 2^N \rightarrow \mathbb{R}$ which corresponds to the \emph{value} of all sub-coalitions $\mathcal{S} \subseteq \mathcal{N}$.
%(i.e., $\nu(\mathcal{S}), \forall \mathcal{S} \subseteq \mathcal{N}\!\setminus \!\{\emptyset, \mathcal{N} \} $).
To overcome  the computational  challenge,  we capitalize on  a constraint generation technique  to gradually approach the \emph{least core} instead of solving problem \eqref{eq:least core} all at once. 
The main idea  contains three steps:  
\emph{i)} solve the relaxed problem \eqref{eq:relaxed least core}  corresponding to a subset of coalitions $\mathcal{F}_1$ (e.g.,  start with $\mathcal{F}_1 = \{\{1\}, \{2\}, \cdots, \{N\}\}$); 
\emph{ii)} identify the most ``violated" sub-coalition (i.e., maximum \emph{excess of coalition}) with the obtained cost allocation; 
\emph{iii)}  add  the identified coalition to $\mathcal{F}_1$.
This process is repeated  until no ``violated"  sub-coalitions  with the obtained cost allocation exists. This indicates the \emph{least core} defined in problem  \eqref{eq:least core} is approached.  We defer the implementation of  constraint generation technique later. 
%(the cost allocation candidates with the minimum maximal \emph{excess of coalition}) by continuing solving a sequence of  \emph{lexicographically optimization} problem.
%
% first start with a relaxed problem as follows and then dynamically approach the \emph{least core} based on the constraint generation procedure \cite{hallefjord1995computing}.

{\small 
	\begin{equation}\label{eq:relaxed least core}
	\begin{split}
	z^{*, 1}= \!=\!\left[
	\begin{array}{l}
	\min z \\
	\text{subject~to:}\\
	\quad   x(\mathcal{N}) = \nu(\mathcal{N}) \\
	\quad   x(\mathcal{S}) -\nu(\mathcal{S}) \leq z, ~\forall \mathcal{S} \in \mathcal{F}_1\!\setminus\! \{\emptyset, \mathcal{N} \}. 
	\end{array}
	\right]
	\end{split}
	\end{equation} }

\subsubsection{Lexicographically optimization}
Intuitively,  if the \emph{least core} is unique, the \emph{nucleolus} is found. However, that is not the usual case  and  we usually have to 
carry on to identify the unique \emph{nucleolus} by solving a sequence of lexicographically optimization problems. For example, minimize $O_2(\bm{x})$ over the \emph{least core }  $\bm{x} \in \{ \bm{x} | O_1(\bm{x}) = z^{1, *} \}$ and so forth. Considering the general case, we introduce the problem of  minimizing  $O_k(\bm{x})$ over $\{ \bm{x}  | O_j(\bm{x}) = z^{j, *}, \forall j= 1, 2, \cdots, k-1 \}$ at episode $k$.
%Similarly, it's computationally intractable to obtain the cost allocation that minimize  $O_k(\bm{x})$  due the the entire information of the characteristic functions. 
%To achieve the objective, we are required to solving  the following   problem: 
%\begin{equation} \label{eq: lexicographically optimization}
%	\begin{split}
%	& z^{k, *} = \min z  \\
%	& \text{subject~to:} \\
%	&\quad \quad  z^{i, *} = \nu(\mathcal{S}) - x(\mathcal{S}),  \forall \mathcal{S} \subseteq \mathcal{\bar{N}}, \forall i = 1, \cdots, k-1. \\
%    &\quad \quad  z \leq \nu(\mathcal{S}) - x(\mathcal{S}),   \forall \mathcal{S} \subseteq \mathcal{N} \setminus  \mathcal{\bar{N}}.\\
%	&\quad \quad  x(\mathcal{N}) = \nu(\mathcal{N}). 
%	\end{split}
%\end{equation}
%where $\mathcal{\bar{N}}$ indicates coalitions corresponding to the \emph{binding} constraints as of step $k-1$.  These constraints can be identified in the process of the lexicographically optimization. However, the underlying challenge to solve problem \eqref{eq: lexicographically optimization} is the requirement of the coalition values corresponding to the inequality constraints which is computationally cumbersome as explained. 
Akin to  computing \emph{least core}, we capitalize on the constraint generation technique to overcome the computation burden by executing  the three steps. 
	% Similarly, instead of solving the problem $O_k(\bm{x})$ over $\{ \bm{x}  | O_j(\bm{x}) = z^{j, *}, \forall j= 1, 2, \cdots, k-1 \}$ exactly all at once with the entire characteristic function,  
	%caused by the entire characteristic function.  
	%we gradually approach the solution 
	%of the lexicographically optimization $\big\{  \min_{\bm{x}} O_k(\bm{x}):  \bm{x} \in \{ \bm{x}  | O_i(\bm{x}) = z^{i, *}, \forall i = 1, 2, \cdots, k-1 \} \big\}$
	% by following the three steps. % updated with the identified most ``violated" coalitions.
	%Particularly, each episode $k$ composed of multiple iterations. 
	%At each iteration, we successively execute the following three procedures: 
	%\emph{i)} solve  relaxed problem \eqref{eq:masterproblem} corresponding to a  subset of coalitions  $\mathcal{F}_k$ (e.g., the grand coalition and the singleton coalitions), 
	Slightly different,  we have the relaxed problem \eqref{eq:masterproblem} with  blocks of \emph{binding} constraints  indicated by $\Gamma_j$. The interpretation  is that 
	at each episode $k$, we solve the lexicographically optimization $\min_{\bm{x}} O_k(\bm{x})$   within the scope of $\{ \bm{x} \in \mathbb{R}^n \vert O_j(\bm{x}) = z^{j, *}, \forall j = 1, 2, \cdots, k-1\}$.
%\emph{ii)} apply the constraint generation technique to identify the most ``violated"  sub-coalitions with the obtained cost allocation;  
%\emph{iii)} add the identified coalition to the sub-coalition $\mathcal{F}_k$ of  problem \eqref{eq:masterproblem}.  We repeat the iterations for  episode $k$ until  no ``violated"  sub-coalition exists.   
%Particularly, if the solution of  problem  \eqref{eq:masterproblem} is unique,  the \emph{nucleolus} is approached otherwise continue. %we have to carry on. 

\vspace{-3mm}
% (P_M^k)~~ 
{\small 
	\begin{equation} \label{eq:masterproblem}
	\begin{split}
	&z^{k, *} \!=\!\left[
	\begin{array}{l}
	\min ~z \\
	%\text{subject~to:}~\\
	\text{subject~to:}~    x(\mathcal{S})-\nu(\mathcal{S}) \leq z, \\
	\quad  \quad \quad \quad  \quad  \forall \mathcal{S} \in \mathcal{F}_k \setminus \cup_{j \leq k-1} \Gamma_j. \\
	\quad   z^{j, *} = x(\mathcal{S})-\nu(\mathcal{S}), \\
	\quad    \quad \quad \quad  \forall \mathcal{S} \in \Gamma_{j}, j= 1, 2, \cdots, k-1.  \\
	\quad   x(\mathcal{N}) = \nu(\mathcal{N}). \\
	\end{array}
	\right]
	\end{split}
	\end{equation}}
% where  $z^{i,*}$ represents the optimal value at step $i$.  $\Gamma_i$ collects the \emph{binding} constraints at step $i$; This enforces   $O_i(\bm{x}) = z^{i, *}, \forall i = 1, 2, \cdots, k-1$. 

% Similar to the starting point of \emph{least core}, if the solution of the \emph{master} problem \eqref{eq:masterproblem} is \emph{unique},  the \emph{nuclenous} is approached, otherwise we continue the lexicographically optimization by moving to the next tep $k+1$.
% 
% we proceed the following step to search for the most ``violated" constraints 
% (most dissatisfied coalitions) regarding the current cost allocation $\bm{x}^{k, *}$.  We refer this step as constraint generation. 
\vspace{-2mm}
\subsubsection{Constraint generation}
This part introduces the implementation of  the constraint generation to identify the most ``violated" sub-coalition  for 1) and 2). 
%while computing the \emph{least core}  and solving the \emph{lexicographically minimization} problems.

%employed in computing the \emph{least core} and \emph{lexicographically optimization} to reduce computation. 

%As claimed, both the computing of \emph{least core} and \emph{lexicographically optimization} depend on the constraint generation step to add constraint corresponding to the locally most violated coalitions. 

% This part introduces the implementation of the constraints generation technique for the CES coalition to reduce computation in the 

Suppose we have a cost allocation  $\bm{x}^{k, *}$ (the solution of problem \eqref{eq:relaxed least core} or \eqref{eq:masterproblem} with a specific subset of  coalitions $\mathcal{F}_k$).  The constraint generation technique requires to identify the 
most ``dissatisfied" or ``violated" sub-coalition  (the subset of players) not included in $\mathcal{F}_k$. This requires to identify the sub-coalition with the maximum  \emph{excess of coalition} regarding the  cost allocation $\bm{x}^{k, *}$  within  the remaining coalitions {\small $\mathcal{N}\setminus\mathcal{F}_k$}. 
% details by starting with an obtained cost allocation proposal $\bm{x}^{k, *}$.
% Actually this step attempts to identify the  critical constraints that determine  \emph{nucleolus} (at most $N$). The underlying idea is  straightforward. 
%Based on the current cost allocation,  
%We can figure out  the  ``most dissatisfied" coalition (subset of players) that not included in $\mathcal{F}_k$ (or $\Omega$)  regarding  the current cost allocation $\bm{x}^{*, k}$ by maximizing 
%the \emph{excess of coalitions}. 
% as the critical constraint for computing the \emph{nucleolus}. With such information in hand,  we can adjust the current cost allocation to reduce the ``dissatisfaction" of the players by adding such constraint in  $\mathcal{F}_k$ ($\mathcal{F}_0$ when computing the \emph{least core} at the beginning).
% 
%In the subsequent, we introduce the constraint generation step to identify the most ``violated" coalitions at each step.
%In particular, this step needs to exclude the coalition already included in  $\mathcal{F}_k$ (or $\Omega$). Therefore, we need to exclude such coalitions to be generated again in this step. 
To address such issue, we define some binary variables $s_i \in \{0, 1\}, \forall i \in \mathcal{N}$ to indicate whether player $i$ is in the identified sub-coalition or not.  
Thus we can interchangeably indicate a coalition by  $\mathcal{S}^j  \subseteq \mathcal{N}$ or a binary vector $\bm{s}^j = \{s^j_1, s_2^j, $ $ \cdots,  s_N^j\}$, where we  have $s_i^j = 1$ if player $i$ is in coalition $\mathcal{S}^j$, otherwise $s_i^j=0$.   
Thus,  we can formulate the problem  as

\vspace{-10pt}
{\small 
	\begin{subequations} \label{eq: Constraint generation}
		\begin{align}
		c^{*} =\max_{\mathcal{S}}  &\sum_{j \in \mathcal{S}} x^{k, *}_j- z^{k, *}-\nu(\mathcal{S})   \notag\\
		%&\text{subject~to:}  \notag \\
		\text{subject~to:} &  \label{eq:8b} 1 \leq \sum_{i \in \mathcal{N}} s_i \leq N- 1.  s_i \in \{0, 1\}, ~\forall i \in \mathcal{N}. \\
		&\label{eq:8c}   \sum_{\mathclap{\{i|s^j_i =0\}}} s_i + \sum_{\mathclap{ \{i|s^j_i=1\}}} (1-s_i) \geq 1,  ~\forall j| \mathcal{S}^j \in \mathcal{F}^k. 
		%& \label{eq:8d}  s_i \in \{0, 1\}, ~\forall i \in \mathcal{N}.  
		\end{align}
\end{subequations} }
where  constraint \eqref{eq:8b} is imposed to exclude the empty coalition $\emptyset$ and  grand coalition $\mathcal{N}$. 
Constraint \eqref{eq:8c} enforces the exclusion of  coalitions $\mathcal{F}_k$.   
% The intuitive interpretation is that  at least one player out of  the coalitions $\mathcal{F}_k$ is included or at least one player in the coalitions $\mathcal{F}_k$ is excluded. 

We note that problem \eqref{eq: Constraint generation} requires the explicit characteristic function $\nu(\mathcal{S})$. 
However, for the ES coalition game,  the  characteristic function $\nu(\mathcal{S})$ is characterized by the stochastic optimization problem \eqref{eq:MainProblem}.  To address such issue, we blend the problem  as  

\vspace{-3mm}
{\small 
	\begin{equation}  \label{eq:constraints_generation}
	\begin{split}
	&c^{*} =  \max   \sum_{j \in \mathcal{S}} x^{k, *}_j - z^{k, *} - \big( c(\bm{x}_{\mathcal{S}})+\sum_{\omega \in \Omega} \rho_\omega g(\bm{x}_{\mathcal{S}}, \zeta_\omega) \big) \\
	%& \text{subject~to:} \\
	&  \text{subject~to:} ~~ \eqref{eq:1b}-\eqref{eq:1d}.  \eqref{eq:4a}-\eqref{eq:4e},~ \eqref{eq:4g}-\eqref{eq:4h},~\forall i \in \mathcal{N}. \\
	& \quad p^{ \text{g+}, \omega}_{i, t}- p^{\text{g-}, \omega}_{i, t} \geq s_i p^{\text{b}, \omega}_{i, t}+p^{\text{d}, \omega}_{i, t}-p^{\text{r}, \omega}_{i, t}, \\
	& \quad p^{ \text{g+}, \omega}_{i, t}, p^{\text{g-}, \omega}_{i, t} \leq s_i P^{g, \max} \\ 
	& \quad 1 \leq \sum_{i \in \mathcal{N}} s_i \leq N- 1.  \quad  s_i \in \{0, 1\}, ~\forall i \in \mathcal{N}.   \\
	&  \quad \sum_{\mathclap{\{i|s^j_i =0\}}} s_i + \sum_{\mathclap{ \{i|s^j_i=1\}}} (1-s_i) \geq 1,  ~\forall j| \mathcal{S}^j \in \mathcal{F}^k.  \\
	%	&\quad  s_i \in \{0, 1\}, ~\forall i \in \mathcal{N}.   \\
	&\quad \text{var:} ~  E_{\cal, S}, P_{\cal S}, \bm{s}, \bm{y}^{\omega}_i, \forall i \in \mathcal{N}. 
	\end{split}
	\end{equation}}
where $P^{g, \max}$ indicates the maximum trading energy with the grid of  each building over single period. 
The objective of problem \eqref{eq:constraints_generation} characterizes the \emph{excess of coalition} for  an ES coalition.  Particularly, we use the combinatorial  constraints  $p^{ \text{g+}, \omega}_{i, t}- p^{\text{g-}, \omega}_{i, t} \geq s_i p^{\text{b}, \omega}_{i, t}+p^{\text{d}, \omega}_{i, t}-p^{\text{r}, \omega}_{i, t}$ to uniformly capture the load balance for  the buildings  in or out of the sub-coalition. 
Specifically, for the buildings in the identified sub-coalition, we  have $s_i=1$ and the procured electricity from the grid  is at least to  satisfy the building demand,  otherwise we have $s_i=0$ and the load balance constraints are relaxed.

%The binary variables $s_i$ can be interpreted as indicator whether the building participant $i$ is included in the identified coalition. 
%If not included, we can assign $s_i =0$ and thus the energy consumption requirement is not accounted in the formulation. 

Problem \eqref{eq:constraints_generation} is a mixed-integer linear programming  (MILP) with $O(N)$ binary variables, which can be handled  by some  existing solvers like CPLEX for moderate scales.  However, if the scale is very large and solving problem \eqref{eq:constraints_generation}   directly becomes computationally intensive,  we would need to find some other ways to handle the problem. 
As aforementioned, with the  (most) ``dissatisfied''  sub-coalition $\mathcal{S}^{*}$ regarding the current cost allocation  proposal $\bm{x}^{k, *}$ obtained, our next step is to added it to 
%Note that by solving problem \eqref{eq:constraints_generation},  the most ``dissatisfied''  sub-coalition $\mathcal{S}^{*}$ with the current cost allocation  proposal $\bm{x}^{k, *}$ %can be identified, which will be added to 
$\mathcal{F}_k$ (i.e., $\mathcal{F}_k: = \mathcal{F}_k \cup \{\mathcal{S}^{*}\}$) and  adjust the cost allocation accordingly.  

The procedures of identifying the most ``violated" sub-coalition and adjusting the cost allocation are  alternated until no ``violated" sub-coalition is found, i.e., the optimal value of problem \eqref{eq: Constraint generation}  is non-positive ($c^{*} \leq 0$).  This implies the optimal solution of problem \eqref{eq:relaxed least core}  or \eqref{eq:masterproblem}  is approached, i.e., 
the \emph{least core} or the cost allocation for $\min O_k (\bm{x})$   has been identified. 

We display the main procedures to search for the \emph{nucleolus} of the ES coalition game in \textbf{Algorithm} \ref{alg:algorithm1}. Particularly,  we clarify three main points regarding the algorithm. 
\emph{First}, the algorithm includes two-loops: \emph{outer-loop} and \emph{inner-loop}.
The \emph{outer-loop} associates with the  lexicographically optimization indicated by  the episode $k$. 
Whereas  the \emph{inner-loop} iteratively solve the lexicographically optimization by employing constraints generation technique.  
In \emph{inner-loop}, we alternatively  identify the most ``violated" coalition and update the cost allocation.
The \emph{inner-loop} will terminate until no ``violated" sub-coalition is found (i.e., $c^{*} \leq 0$ for problem \eqref{eq:constraints_generation}), which indicates the  lexicographically optimization  has been solved.   \emph{Second}, there are  two crucial steps when switching from the \emph{inner-loop} to  the \emph{outer-loop}: \emph{i)} at the end of  each \emph{inner-loop}, the \emph{active} or \emph{binding} constraints  are required to be identified (line 12). This can be achieved by checking the inequality constraints  of  problem \eqref{eq:relaxed least core}  or \eqref{eq:masterproblem}. % corresponding to the coalitions% \mathcal{F}_k \setminus \cup_{i \leq k-1} \Gamma_i.
Specifically, with the obtained solution $\bm{x}^{k, *}, z^{k, *}$,  we identify the \emph{active} or \emph{binding} constraints  $\Gamma_k$ for next computing epoch by checking the equality $ z^{k, *} = \nu(\mathcal{S})-\bm{x}(\mathcal{S}),$ $\forall \mathcal{S} \in \mathcal{F}_k \setminus \cup_{i \leq k-1} \Gamma_i$;   \emph{ii)}  the subset of coalitions is copied for the next computing epoch $k+1$ (line 13). 
\emph{Third}, the overall algorithm will terminate until the solution of lexicographically optimization is unique (line  4-7). 
%
%$\forall \mathcal{S} \in \mathcal{F}_k \setminus \cup_{i \leq k-1} \Gamma_i$ with  $ z^{k, *} = \nu(\mathcal{S})-\bm{x}(\mathcal{S})$, we have $\mathcal{S} \in  \Gamma_k$. 
%Besides, we note the \emph{out-loop} will terminate until an unique solution for a \emph{lexicographically optimization}  is approached. 

\begin{algorithm}
	\caption{Search for  nucleolus of ES coalition game based on constraints generation} \label{alg:algorithm1}
	\SetAlgoLined
	\Input{$\mathcal{N}:=\{1, 2, \cdots N\}$: building participants. }
	\Output{$\bm{x}  \in \mathbb{R}^N$: cost allocation  for the buildings.}
	\textbf{Initialize:} $k \rightarrow 1$, $\mathcal{F}_1 =\{ \{1\}, \cdots, \{N\} \}$, \textbf{ \small STOP}: = false\;
	\While{!\textbf{\small STOP} }{
		Solve  prolem \eqref{eq:masterproblem}  (or \eqref{eq:relaxed least core} if $k=1$) and obtain the solution $\bm{x}^{k, *}$ and $z^{k, *}$\;
		\If{ the solution is unique}{
			\textbf{\small STOP}: = true\;
			\textbf{\small break}\;
		}
		Solve problem \eqref{eq:constraints_generation} to  identify  the most ``dissatisfied" sub-coalition $\mathcal{S}^{*}$ and the corresponding \emph{excess of coalition} $c^{*, k}$ with the current cost allocation $\bm{x}^{k, *}$\;
		\eIf{$c^{k, *} > 0 $}{
			%\textbf{ \small STOP}: = true;\; 
			Add the identified sub-coalition: $\mathcal{F}_{k}: = \mathcal{F}_k \cup\{\mathcal{S}^{*}\}$\; % \Comment{add cuts}\;}
			
		}{
			Identify the \emph{active and binding}  constraints $\Gamma_k$\;
			$\mathcal{F}_{k+1}:  = \mathcal{F}_k$;\\	
			$k: = k+1$\;
		}
	}
	
\end{algorithm}

\section{Case Study}
This section reports the numeric results. 
We first study the fairness  and computational efficiency of the cost allocation based on \emph{nucleolus}.  We then investigate the enhanced  economic benefits of the ES sharing model  over the IES model.   %The simulation setup refers to \cite{yang2020optimal}.

\subsection{Simulation setup}
We set up  the case studies  based  on the  real  building demand profiles \cite{BuildingDemand} and renewable generation profiles (i.e., wind and solar power) \cite{WindSolar} for one year (i.e., 365 scenarios).  To account for the complementary feature of energy use in buildings,  multiple types of buildings (e.g., office, hotel, school, hospital and restaurant) are considered. 
We display the typical demand  and renewable generation profile  in Fig. \ref{Fig: BuildingDemand}. 
Considering the large number of scenarios lead to high computation cost,  we  choose  $S=10$ representative scenarios to capture the patterns of renewable generations and building demands, respectively. 
The  ES charging and discharging  efficiency  is set as $\eta^{\text{ch}}, \eta^{\text{dis}} = 0.9$.  For the amortized ES capital price, we assume an annual interest rate   $r= 0.06$ and  ES lifetime $L=10$ years.
We study  the problem on a daily circle with the time equally  discretized into $T=24$ time slots, corresponding to a decision interval of $\Delta = 1$h. 
We refer to the time-of-use electricity price of Singapore: 
$c^{\text{g+}}_t = 0.1271${\small s\$\si{\kilo\watt\hour}} (off-peak 23:00-7:00) and $c^{\text{g+}}_t=0.2085${\small s\$\si{\kilo\watt\hour}}(peak 8:00-22:00) and demand charge  $c^{g, \max}= 0. 1335${\small s\$\si{\kilo\watt}} (we set selling price as $c^{\text{g-}}_t =0$).  The maximum trading power  with the grid is set as  $P^{g, \max} = 10^3$ {\small \si{\kilo\watt}} for each building. 

%For the estimation of the battery ES price model, we set the annual interest rate as   $R= 0.06$ and the standard lifetime as $L=10$.
\vspace{-15pt}
\subsection{Fairness and computational efficiency}
This part  evaluates the fairness and computational efficiency of the cost allocation  for ES sharing (\textbf{Algorithm} \ref{alg:algorithm1}).
We consider five ES coalition of different scales: $N:=\{3, 5, 8, 10, 20\}$.   
%To illustrate the fairness of the cost allocation as well as the computation efficiency, 
We compare the cost allocation  based on \emph{nucleolus} with 
\emph{proportional method} \cite{liu126optimal} and \emph{Shapley approach}  \cite{saad2009coalitional}. 
The \emph{proportional method} is empirical and  easy to compute   
whereas  \emph{Shapley approach} is  more sophisticated  but computationally intensive.

\begin{itemize}
\item \emph{ Proportional method}:  This method  distributes the ES capital cost among the buildings  based on their proportions of operation cost (electricity bill) reduction relative to  no ES case (\emph{benchmark}). 
 %(the operation cost of each building is readily available through the net-metering).  
 Specifically, we denote the operation cost of the buildings with no ES as $\bar{\bm{x}}^{\text{Opex}} \in \mathbb{R}^N$ (this is readily available as the buildings have to purchase the deficiency instantaneously  in such case). Besides, we can obtain the operation cost $\bm{x}^{\text{Opex}} \in \mathbb{R}^N$ and the ES capital  cost $y^{\text{Cap}} \in \mathbb{R}$ with the ES sharing model  by solving problem \eqref{eq:MainProblem}.   As the operation cost of  each  building is readily  available through the net metering,   the \emph{proportional method}  only distributes  the ES capital cost of ES sharing model according to 
 
 {\small
\[
   x^{\text{Cap}}_i =  (\bar{x}^{\text{Opex}}_i-x^{\text{Opex}}_i)\frac{y^{\text{Cap}}}{\bm{1}^T (\bar{\bm{x}}^{\text{Opex}} - \bm{x}^{\text{Opex}} )}, ~\forall i \in \mathcal{N}. 
\] }
where $\bm{1} \in \mathbb{R}^N$ denotes a $N$-dimensional unit vector. 

Thus, the total cost for each building  amounts to $x^{\text{Opex}}_i + x^{\text{Cap}}_i$ ($\forall i \in \mathcal{N}$). 

\item  \emph{Shapley approach}: The underlying idea of \emph{Shapley approach} is to distribute  payoff  among the players  based on  their marginal contributions to ES coalition.  Specifically, for a coalition game $(\mathcal{N}, \nu)$, the payoff assigned to each player is calculated as  \cite{saad2009coalitional}  

{\small
\begin{equation*} \label{eq: ShapleyValue}
\begin{split}
x_i(\mathcal{N}) = \sum_{\mathclap{\mathcal{S} \subseteq \mathcal{N}\setminus \{i\} }}\frac{\vert \mathcal{S} \vert! ( \vert \mathcal{N} \vert -\vert \mathcal{S} \vert -1)!}{\vert \mathcal{N} \vert !} [\nu(\mathcal{S} \cup \{i\}) - \nu(\mathcal{S}) ] 
\end{split}
\end{equation*}}
where $\mathcal{S}$ denotes all subsets of players of $\mathcal{N}$ that exclude player $i$. 
%The underlying idea of Shapley value can be interpreted as follows. 
%We note that the allocated cost to each player $i$ is built by the 
Clearly,  the  allocated payoff for  player $i$ is built by its  marginal  contribution to each coalition $S$: $\nu(\mathcal{S} \cup \{i\}) - \nu(\mathcal{S})$.
%
%marginal ``contribution" of player $i$ to an existed coalition $S$ is quantified by $\nu(\mathcal{S} \cup \{i\}) - \nu(\mathcal{S})$. Correspondingly, the weight denotes the probability that player $i$ faces coalition $\mathcal{S}$ when joining in a random order (i.e., the collection of players join in front of $i$ is $\mathcal{S}$). 
\end{itemize}

%The\emph{proportional allocation} is an  empirical  cost allocation that  is  easy to compute;   
%whereas  \emph{Shapley value} method can provide a \emph{fair} cost allocation  but is computationally intensive as it requires the entire information of the characteristic functions.
% \emph{Shapley value} method is computationally intensive as the value of all coalitions $2^N-1$ is required as indicated in \eqref{eq: ShapleyValue}.

As aforementioned, we evaluate a cost allocation to be fair  if all the players  are satisfied.
In this paper, we assume all the building players  are profit-oriented and have no other preferences, therefore 
	it is reasonable to quantify their  \emph{dissatisfaction} (\emph{satisfaction})  by  their  allocated cost.  %In this paper, we use the \emph{satisfaction} of players to quantify the \emph{fairness} of a cost allocation.
To account for the group of players,   we focus on the minimum \emph{excess of coalition} which  can be used as an indicator of  minimum \emph{dissatisfaction} (DSAT) over all the buildings regarding the cost allocation. Specifically, we have

%\emph{satisfaction} (SAT) of all  players as an indicator of \emph{fairness}, which can be quantified by the minimum \emph{excess of coalition} with the cost allocation $\bm{x}$: 
%
%Before we present the main results, we first define the  \emph{satisfaction} (STA) metric, i.e., the minimum \emph{excess of coalition} for a specific cost allocation $\bm{x}$ as an indicator of \emph{fairness} in this paper.
\begin{equation}
\text{DSAT}= \min_{{\mathcal{S} \subseteq \mathcal{N},\\
		\mathcal{S} \neq \emptyset, \mathcal{N}} } e(\bm{x}, \mathcal{S}) %\bm{x}(\mathcal{S})-\nu(\mathcal{S}) 
\end{equation}
Clearly,  all players are satisfied (i.e., fair)  if the minimum \emph{excess of coalition} is non-positive (DSAT $\leq 0$).  Moreover, we prefer a cost allocation with a more negative DSAT.

For each scale, we apply the  three cost allocation methods to  achieve the \emph{ex-post} cost allocation across the building participants. 
	For notation, the buildings are labeled by B1-B20 with the  allocated  cost  displayed 
	in TABLE \ref{tab:N3} (the results for  $N=20$ are omitted due to space limits).  
	First of all, we note that the total cost for each scale are the same regardless of the \emph{ex-post} cost allocation mechanism used.   
	This is caused by the same optimization problem \eqref{eq:MainProblem} we rely on to compute both the \emph{Shalepy} and \emph{propertional} allocations. 
	However, there exist some differentials regarding the allocated cost to each building under the different cost allocation mechanisms, which  lead to the different DSAT as indicated in the last row of each table. 
Particularly,  we find  that for the scales $N:=\{3,5\}$, all the three cost allocation mechanisms   can ensure fairness  as indicated by the  negative DSAT,   whereas  for the larger scales $N:=\{8, 10\}$, the fairness is only  ensured by the proposed  method and \emph{Shapley approach}  not the \emph{proportional method}.  
This demonstrate that  the proposed method and  \emph{Shapley approach} can  ensure  \emph{fairness}  whereas  \emph{proportional method} may fail in some cases. 

For the computational efficiency, we quantify the computation cost  of different methods by  the fraction of characteristic function required  (i.e., the number of coalition \emph{value} computed by solving problem  \eqref{eq:MainProblem}). 
The computation cost for  different scales  ($N:=\{3, 5, 8, 10, 20\}$) are presented in TABLE \ref{tab:computation}.
Particularly, for the  \emph{nucleolus}, we start with the singleton and grand coalitions (i.e., $\mathcal{F}_1$), and we record  the number of constraint generations (i.e., $\bar{K}$) performed in the execution.
First of all,  we observe that  \emph{Shapley approach} shows the highest computation cost with the entire characteristic function ($2^N-1$ coalition \emph{value}) required.   On the contrary,  \emph{proportional method} is   most efficient and only requires $N+1$ coalition \emph{value} to achieve the cost allocation ($N$ corresponds to computing the cost for each building with no ES and $1$ corresponds to computing the total cost of  grand coalition).
Notably, we observe the computation cost with the  \emph{nucleolus}   is slightly higher than the \emph{proportional method} but significantly lower than the \emph{Shapley approach}. 
%When we start with $\mathcal{F}_1$ with the singleton 
%Notably, the proposed method based on \emph{nucleolus} also takes  approximately linear computation w.r.t. the scale $N$,  but with a minor increase compared with the \emph{proportional method}. 
%We note that the computation cost of the proposed method is jointly determined by the scale $N$ and the total number of generated constraints  $\bar{K}$ in the  execution. 
%Comparatively, the proposed method based on \emph{nucleolus}
% is comparative with the \emph{proportional allocation}  in the computation cost, which both present a linear increase rate w.r.t the scale $N$. 
%  Clearly, the total computation cost for the proposed method is determined by the number of constraint generation steps before the unique \emph{nucleolus} is encountered in the process of lexicographically optimization.  
It's noteworthy that for $N=10$,  only $2.54\%$ (26/1023) of  the  characteristic function  is required,  and when the scale is increased  to $N=20$, the computation burden is reduced to less than 1\% $(88/10^7)$. 
This demonstrates the superior computational efficiency of  \emph{nucleolus} over the \emph{Shapley approach}. 
Therefore,  we conclude that the \emph{nucleolus} outperforms \emph{Shapley approach}  and  \emph{proportional method }  by providing both computation efficiency  and fairness.

%Specifically, benchmarked on the Shapley value method, one $2.54\%$ computation cost is required to achieve the \emph{fair} cost allocation in the CES coalition game. 

%We note that both the proportional allocation and the nucleolus are computationally effective with approximate linear increasing computation burden w.r.t. the buildings whereas the computation burden of the Shapley value based method is exponentially increasing w.r.t. the number of buildings. 
%
%Therefore, by comparing the three cost allocation mechanisms both from the fairness and computational efficacy, we imply that the 
%cost allocation based on the \emph{nucleolus} is suitable suitable for application.

\vspace{3mm}
\begin{table}[h]
	\begin{center}
		\caption{Cost allocations of different methods}
		\label{tab:N3}
		\begin{tabular}{c|c|c|c|c}
			\toprule % <-- Toprule here
			%\multicolumn{4}{c}{$N=3$} \\
			%\hline
			Scale & \multirow{2}{*}{\textbf{ Build.} }& \textbf{Proportional} & \textbf{Shapley} & \textbf{Nucleolus}\\
			(N) &  &  $\times 10^2$(s\$)      &  $\times 10^2$ (s\$)  &  $\times 10^2$ (s\$) \\
			\midrule % <-- Midrule here
			\multirow{4}{*}{3}  & B1  & 2.63 & 2.64   &    2.69\\
			& B2  & 4.46 & 4.49  &    4.44\\
			& B3  & 2.40  & 2.37   &   2.38\\
			& DSAT  & -5.58(Y) & -6.37(Y)     & -5.80(Y)     \\
			\bottomrule % <-- Bottomrule here
		\end{tabular}
		~\\
		~\\
		~\\
		\begin{tabular}{c|c|c|c|c}
			\toprule % <-- Toprule here
			%\multicolumn{4}{c}{$N=5$} \\
			%\hline
			Scale &\multirow{2}{*}{\textbf{ Build.}} & \textbf{Proportional} &  \textbf{Shapley} & \textbf{Nucleolus}\\
			(N) &  & $\times 10^2$(s\$)   &   $\times 10^2$(s\$)       &  $\times 10^2$ (s\$) \\
			\midrule % <-- Midrule here
			\multirow{6}{*}{5} & B1   &  2.57 & 2.57   &  2.59   \\
			& B2   &   4.43 & 4.43   &    4.48 \\
			& B3   & 2.30  & 2.27   &    2.26 \\
			& B4   &  6.66 & 6.72   &     6.75\\
			& B5   &    4.72 & 4.68  &   4.60\\
			& DSAT & -2.00(Y) & -6.73(Y)    & -10.18(Y)     \\
			\bottomrule % <-- Bottomrule here
		\end{tabular}
		~\\
		~\\
		~\\
		\begin{tabular}{c|c|c|c|c}
			\toprule % <-- Toprule here
			%\multicolumn{4}{c}{$N=8$} \\
			%\hline
			Scale &  \multirow{2}{*}{ \textbf{ Build.} }& \textbf{Proportional} & \textbf{Shapley} & \textbf{Nucleolus}\\
			(N) &  &   $\times 10^2$(s\$)   &  $\times 10^2$ (s\$)     &  $\times 10^2$ (s\$) \\
			\midrule % <-- Midrule here
			\multirow{9}{*}{8}  & B1    &  2.61   & 2.59     &  2.64 \\
			& B2    &  4.44   &  4.40    &  4.43 \\
			& B3    &   2.32  &  2.28    & 2.27  \\
			& B4    & 6.59     &  6.62    & 6.60  \\
			& B5    &  4.66   &  4.64    & 4.61  \\
			& B6    &  5.82   &  5.79    & 5.77  \\
			& B7    &  7.54  &  7.56    & 7.54  \\
			& B8    &  6.69  &  6.79    & 6.82  \\
			& DSAT  & 8.34(N) & -0.64(Y)       &-4.98(Y)     \\
			\bottomrule % <-- Bottomrule here
		\end{tabular}
		~\\
		~\\
		~\\
		\begin{tabular}{c|c|c|c|c}
			\toprule % <-- Toprule here
			%\multicolumn{4}{c}{$N=10$} \\
			%\hline
			Scale &  { \textbf{ Build.} }& \textbf{Proportional} & \textbf{Shapley} & \textbf{Nucleolus}\\
			(N) & &   $\times 10^2$(s\$)   &  $\times 10^2$ (s\$)     &  $\times 10^2$ (s\$) \\
			\midrule % <-- Midrule here
			\multirow{11}{*}{10} & B1    &  2.62   & 2.58     &  2.61 \\
			& B2    &  4.44   &  4.44    &  4.43 \\
			& B3    &   2.22  &  2.30    & 2.22  \\
			& B4    & 6.68     &  6.59    & 6.64  \\
			& B5    &  4.60   &  4.68    & 4.59  \\
			& B6    &  5.74   &  5.79    & 5.75  \\
			& B7    &  7.51     &  7.48    & 7.50  \\
			& B8    &   6.77     &  6.71         &  6.82\\
			& B9     &  6.44      &  6.45            &  6.46  \\
			& B10   &  7.97      &    7.96           &    7.97\\
			& DSAT   &      10.49(N)         &       -1.48(Y)        &   -4.85(Y) \\
			\bottomrule % <-- Bottomrule here
		\end{tabular}
		\begin{tabular}{cc}
			Note: Y: satisfied &  N: not satisfied
		\end{tabular}
	\end{center}
\end{table}

\begin{table}[h]
	\begin{center}
		\caption{Computation of different methods}
		\label{tab:computation}
		\begin{tabular}{c|c|c|c}
			\toprule % <-- Toprule here
			\textbf{ Scale }  & \textbf{Proportional}   & \textbf{Shapley} & \textbf{Nucleolus}\\
			(N) &   \textbf{Computation}   &   \textbf{Computation} &  \textbf{Computation}  \\
			\midrule % <-- Midrule here
			3      & 4         & 7              & 8 (4)\\
			5      & 6         & 31            & 13 (7)  \\
			8      & 9        & 255           & 24 (15)\\
			10    &11       & 1023         & 37 (26)  \\
			20  &  21            &    $>10^7$                 &  88     (67)         \\
			$N$  & $N+1$ & $2^N-1$  & $N+1+\bar{K} (\bar{K})$ \\
			\bottomrule % <-- Bottomrule here
		\end{tabular}
	\end{center}
	\vspace{-6mm}
\end{table}

\vspace{-0mm}
\subsection{Economic benefits of ES sharing}
\begin{figure}[h]
	\centering
	\includegraphics[width=3.2 in ]{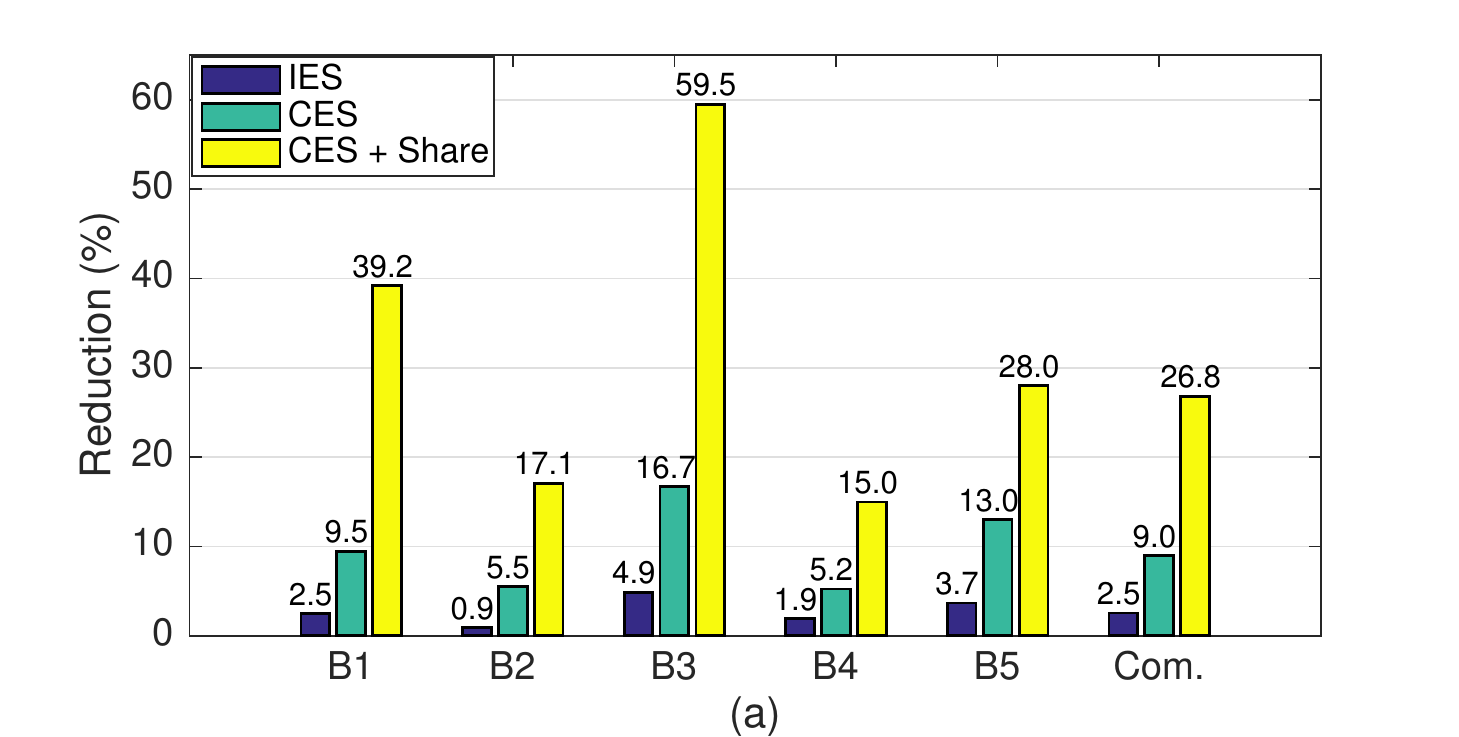}\\
	\includegraphics[width=3.2 in ]{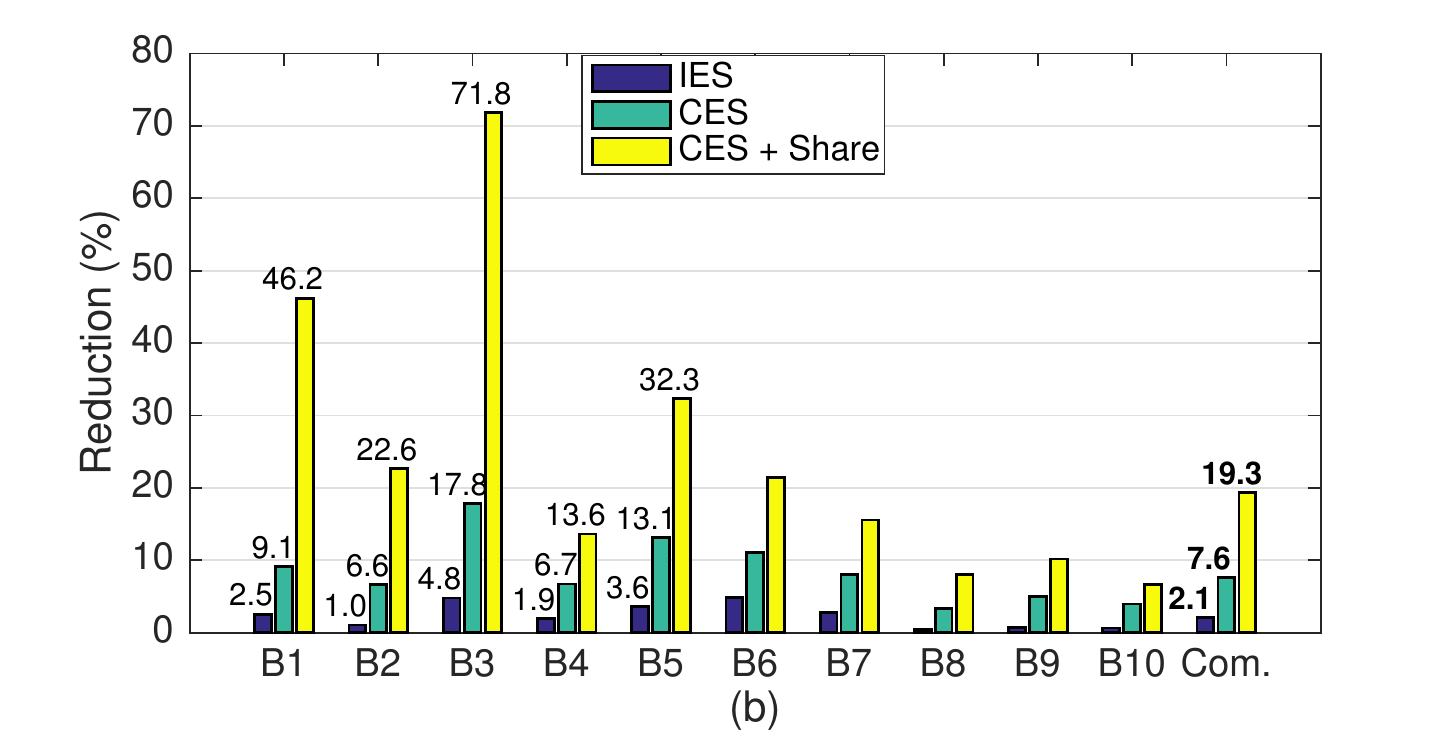}\\
	\caption{The building-wise (B1-B10) and community-wise (Com.) cost reduction  with the IES and CES model: (a) $N=5$. (b) $N=10$ (no ES as \emph{baseline}).}\label{Fig: IES_CES}
\end{figure}

\vspace{-1mm}
In this part, we study the enhanced economic benefits of the ES sharing model  (referred to CES model) over the  IES model. 
For the IES model, each building invests private ES separately where the optimal ES sizing  and operation are obtained by solving problem \eqref{eq:MainProblem}  with $N=1$. 
%Particularly,  to quantify the economic benefits of different ES models,  we use  no ES case as  \emph{benchmark}.  
%To ensure fair comparisons,  we adopt the same data  (i.e., renewable generation) and  parameters (i.e., ES capital price, roundtrip efficiency) in the two models. 
%As an instance,  we investigate a community microgrid composed of $N=10$ buildings. These buildings commit to invest  a common-owned CES and share the \emph{ex-post} cost  according to the proposed cost allocation mechanism. 
%First of all, we investigate the economic benefits of each individual building with the different ES models. 
Particularly, for the CES model, we consider two  settings:  without energy sharing (CES)  and  with energy sharing (CES + Share). 
	For the CES + Share model, we can follow the previous notations and formulations but  replace   \eqref{eq:4d}  with  $\sum_{i=1}^N e^{\text{b}, \omega}_{i, t}  \geq 0, \forall  t \in \mathcal{T}$ which indicates the  stored energy injected by the different buildings are commonly owned.  
	%For the CES model,  \blue{we first solve problem \eqref{eq:MainProblem} to decide the optimal ES sizing and operation of the shared ES.  After that we employ the cost allocation mechanism based on 
	%	\emph{nucleolus} to achieve the \emph{ex-post} cost allocation.} 
	Using the scale with $N=5$ and $N=10$ buildings as examples,  we study both the building-wise (B1-B10)  and community-wise (Com.) economic benefits with the two ES models. The building-wise economic benefits with the CES and CES + Share model are calculated based on  the \emph{ex-post} cost allocation of  \emph{nucleolus}. The community-wise economic benefits represent the overall economic benefits for all the buildings.
	Using the cost with no ES as \emph{baseline}, the 
	building-wise  (B1-B10)  and community-wise (Com.)  economic benefits can be quantified by the cost reduction as shown  in  Fig. \ref{Fig: IES_CES}.  
	%\blue{using no ES as \emph{baseline}, we compare the cost reduction for the buildings  (B1-B10) with  IES and CES model in  Fig. \ref{Fig: IES_CES}. }
	%Specifically, the cost of each building  is stacked by 
	%the ES capital cost (CAPex), the operation cost (Opex) as well as  cost reduction (Reduction)  compared with  no ES case (\emph{benchmark}).
	%Whereby we quantify the cost reduction of each building by using the no ES case as \emph{benchmark}.  ()
	We see  the CES model yields higher percentage of  cost reduction to each committed building  and the whole community over the IES model. 
	Taking the case with $N=5$ as an example ((Fig. \ref{Fig: IES_CES}(a))),  the cost of B3 (i.e., electricity bill plus ES capital cost) is  cut off by 16.7\%  with the CES model  versus 4.9\% with IES model,  
	and the overall cost  is reduced by 9.0\% versus  2.5\%.  
	This implies the CES model can enhance  both the building-wise and community-wise economic benefits over the IES model. 
	Besides, we note that the CES + Share model  can  enhance the economic benefits significantly further.  
	For example, the cost reduction for B3 is up to  59.5\%  with the CES + Share model. 
	%We note that  the buildings can obtain significant revenue through energy sharing as we need a large percentage of cost reductions for all the buildings. 
	We see the similar results with the scale $N=10$. 
	Further, by comparing the results with $N=5$ and $N=10$, we find that B1-B5 (appear in both scales) all  gain higher cost reduction  with the larger coalition (i.e.,$N=10$) (the marginal decrease of  B1 is caused by computing accuracy).  This demonstrates that by forming a  large ES sharing coalition, the economic benefits of the building participants can be further enhance, however this may  require a more powerful central coordinator for coordination.

%\vspace{-5pt}
%\begin{figure}[h]
%	\centering
%	\includegraphics[width=2.8 in, height=1.0 in ]{System_IES_CES.eps}\\
%	\caption{The total cost of the community microgrid with the IES and CES model (no ES as \emph{benchmark}).}\label{Fig: reduction}
%\end{figure}

%\vspace{-5pt}
\begin{figure}[h]
	\centering
	\includegraphics[width=3.2 in, height= 1.6 in ]{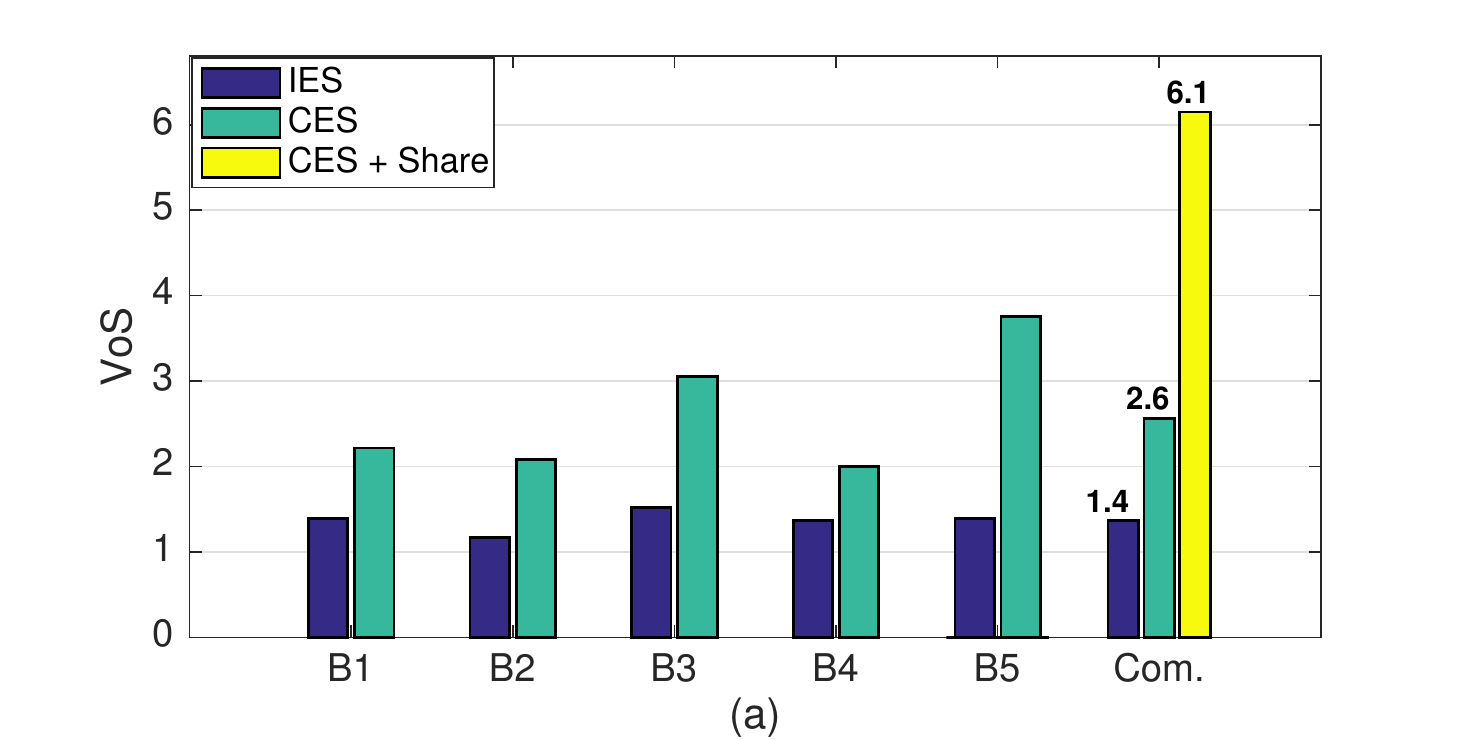}\\
	\includegraphics[width=3.2 in, height= 1.6 in ]{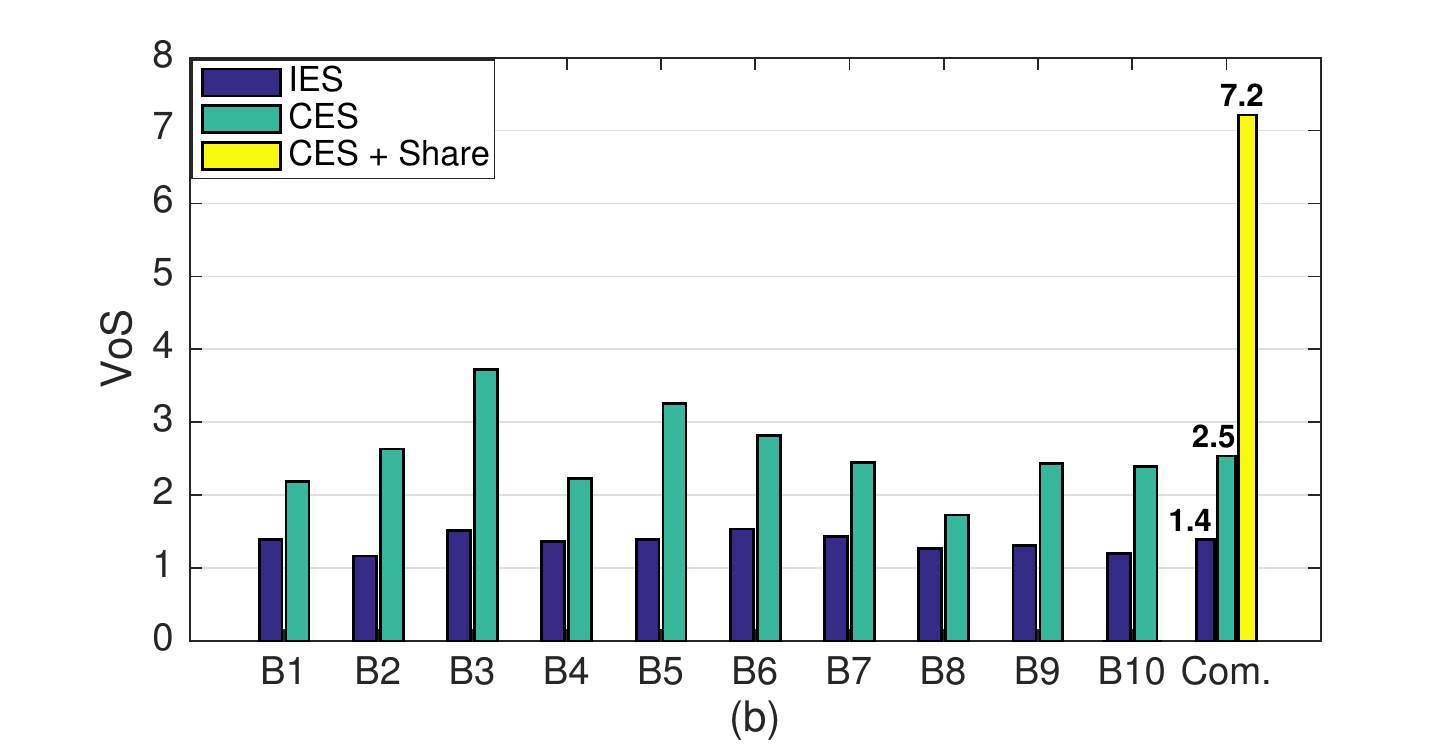}\\
	\caption{The building-wise (B1-B10) and community-wise (Com.) VoS  with the IES and CES model: (a) $N=5$. (b) $N=10$.}\label{Fig: VOS}
\end{figure}

Further, we  study   the  average \emph{value} of ES (VoS) with the different ES models  (i.e., IES, CES, CES + Share).  The VoS  is defined as  the  proportion of   operation cost reduction over the ES capital cost, representing the average return on investment (ROI):
%We define the VoS as the average 
\[  \text{VoS} =[ {\bar{x}^{\text{Opex}} - x^{\text{Opex}}}]/{x^{\text{Cap}}} \]
where $x^{\text{Opex}}$ and $x^{\text{Cap}}$ denote the optimal operation cost and capital cost yield by  an ES model. $\bar{x}^{\text{Opex}}$ represents the operation cost with no ES.  
Intuitively, there exists potential to invest ES if and only if VoS $>1$. 
Moreover, we would prefer an ES model with a higher VoS. 
In this part,  we study the VoS for both individual buildings and the whole community. 
Particularly, the ES capital cost allocated to  each building with the CES model can be obtained by the total allocated cost minus the electricity bill (obtained by  solving problem \eqref{eq:MainProblem}).
For the CES + Share model, we only study the VoS for the whole  community due to the lack of  ES capital cost for individual buildings. 
As with  CES + Share model, the cost of each building consists of the electricity bill, energy trading cost, and ES capital cost. However, the latter two parts can not be distinguished from the cost allocation. 
%We investigate  the VoS for each individual building as well as  average VoS.
Similarly, we use the case with $N=5$ and $N=10$ as examples and we present the results in Fig. \ref{Fig: VOS}.
First, we  observe  that  the  VoS for each building is apparently increased  with the CES  model over the IES model.  
Overall, the VoS for the whole community is about 1.8 times with the CES model than the IES model. 
%This demonstrates the improved ES efficiency with the CES model.  
Besides, by comparing the results with $N=5$ and $N=10$, we find that  a large ES sharing coalition  also favors  the VoS for individual buildings (B1-B5).
Notably, we see the CES + Share model can significantly increase the overall  VoS for the community than the CES model. 
This is reasonable  that  allowing the  buildings to share their  surplus local renewable generation can reduce  the over  grid purchase for supplying the building demand.

%\begin{remark}
%	The CES can help reduce the total cost compared with IES. 
%\end{remark}

%\begin{remark}
%	The CES can increase the average value of storage vs. IES.
%\end{remark}

%We define the following metric to evaluate the average \emph{value} of storage

%Then we have 
%$VoS(CES)=\frac{(21.37-16.18)\times 10^3}{2459} =2.11$ and $VoS(IES)=\frac{(21.37-18.20)\times10^3}{2156}=1.47$

%For the propotional allocation, when $M=8$, it's unstable. 
%55(\{1, 2, 3, 5 ,6\}, -8.33), 63(\{1, 2, 3, 4, 5, 6\}, -3.45), 119(\{1, 2, 3, 5, 6 ,7\}, -5.63), 127( \{1, 2, 4, 5, 6, 7\}, -6.21).
\vspace{-5mm}
\section{Conclusion}
\vspace{-1mm}
	This paper studied a cooperative  energy storage (ES)  business model based on the sharing mechanism.  To maximize the economic benefits of ES, we studied the problem  by integrating the optimal planning (i.e., ES sizing),  operation, and fair \emph{ex-post} cost allocation via  a coalition game formation, thus yielding higher economic benefits to each building and the whole community  over the individual energy storage (IES) model.
	Particularly, the fair \emph{ex-post} cost allocation was achieved based on \emph{nucleolus} which ensures fairness by minimizing the dissatisfaction of all players.
	To handle the exponential computation burden, we applied the constraint generation technique to gradually approach the unique \emph{nucleolus}  considering  the sparse problem structure, which was demonstrated with both computation efficiency and fairness. 
	Further, through the case studies, we found that by enabling energy sharing  through the shared ES, the economic benefits of ES can be further enhanced  for the buildings with surplus local renewable generation. 
	As the commercial deployment of ES is currently impeded by the high capital cost, this work can work as an example how business model designs can benefit the practice of  ES technologies.  Currently, we do not consider the degradation of ES capacity  caused by the charging and discharging circles due to the complexity of quantification, however it seems an interesting  work to incorporate  the recently developed convex  rainflow cycle-based model ~\cite{shi2018optimal}  to address that issue  in the future.

\appendices

%\section{Amortized ES capital price}
%The amortized ES capital price is  calculated as \cite{harsha2014optimal, zhao2019virtual}
%		\begin{displaymath}
%		\begin{split}
%		& k_p = K_p \frac{(1+r)^L}{L \cdot D} ~\text{and}~k_e = K_e \frac{(1+r)^L}{L \cdot D}\\
%		\end{split}
%		\end{displaymath}
%		where $r$ is the annual interest rate.  $L$ is the nominal lifetime of the ES in years (typically 10-20 years).  $D=365$ represents the number of days per year. 
%		$K_p$({\small in s\$\si{\per{\kilo\watt\hour}}})  and $K_E$({\small in s\$\si{\per{\kilo\watt}}}) denotes the ES price for energy and power capacity. %is the price of energy capacity per unit. 
%%		Typically, the lifetime of an ES is $10$-$20$ years. Thus, by using such amortized price model, we project the capital cost of ES on each single day. 
		
		\section{Proof of Proposition 1}
		\begin{proof}
			
			\emph{i)} We first prove constraints $p^{\text{g+}, \omega}_{i,t} p^{\text{g-}, \omega}_{i,t}=0$ is redundant. This is intuitive as $p^{\text{g+},  \omega}_{i, t}, p^{\text{g-}, \omega}_{i, t}>0$ won't happen simultaneously  when the purchase price is higher than  the selling price and the objective is to minimize the total cost. 
			
			\emph{ii)} We prove  constraints $p^{\text{ch}, \omega}_{i,t} p^{\text{dis}, \omega}_{i,t}=0$ can be discarded without affecting the optimal solution by investigating two possible cases. 
			
			\emph{Case I}:   we assume an optimal operation with $p^{\text{ch}, \omega}_{i, t} > p^{\text{dis}, \omega}_{i, t} >0$.
			Since $\eta^{\text{ch}}, \eta^{\text{dis}}<1$, we can construct  another feasible ES operation strategy:
			$\tilde{p}^{\text{ch}, \omega}_{i, t} =p^{\text{ch}, \omega}_{i, t} - p^{\text{dis}, \omega}_{i, t}/(\eta^{\text{ch}} \eta^{\text{dis}})   $ and $\tilde{p}^{\text{dis}, \omega}_{i, t} =0$. It's easy to verify that 
			$\tilde{p}^{\text{b}, \omega}_{i, t} =\tilde{p}^{\text{ch}, \omega}_{i, t} \eta^{\text{ch}} - \tilde{p}^{\text{dis}, \omega}_{i, t}/\eta^{\text{dis}}$ and $\tilde{e}^{\text{b}, \omega}_{i, t+1}= e^{\text{b}, \omega}_{t+1}$ (change the ES operation at time $t$ but do not affect the energy state). 
			Consequently, we would have $\tilde{p}^{\omega}_{i, t} = \tilde{p}^{\text{ch}, \omega}_{i, t} -\tilde{p}^{\text{dis}, \omega}_{i, t} =p^{\text{ch}, \omega}_{i, t} - p^{\text{dis}, \omega}_{i, t}/(\eta^{\text{ch}} \eta^{\text{dis}}) < p^{\text{ch}, \omega}_{i, t} - p^{\text{dis}, \omega}_{i, t}  $ and 
			$\tilde{p}^{\text{g+}, \omega}_{i, t} - \tilde{p}^{\text{g-}, \omega}_{i, t} =\tilde{p}^{\omega}_{i, t} + p^{\text{d}, \omega}_{i, t} - p^{\text{r}, \omega}_{i, t} < p^{\omega}_{i, t} +p^{\text{d}, \omega}_{i, t} - p^{\text{r}, \omega}_{i, t}  = p^{\text{g+}, \omega}_{i, t}-p^{\text{g-}, \omega}_{i, t} $. Therefore, for the constructed operation, we have $\tilde{p}^{\text{g+}, \omega}_{i, t}  < p^{\text{g+}, \omega}_{i, t} $ or $\tilde{p}^{\text{g-}, \omega}_{i, t}  > p^{\text{g-}, \omega}_{i, t} $.
			This implies the constructed solution will yield lower cost,  which contradicts the assumption. 
			
			\emph{Case II}: Assume we have an optimal operation with $p^{\text{dis}, \omega}_{i, t} > p^{\text{ch}, \omega}_{i, t} >0$.  Similarly, we can construct  another feasible ES operation strategy:  $\tilde{p}^{\text{ch}, \omega}_{i, t} =0$ and $\tilde{p}^{\text{dis}, \omega}_{i, t}=p^{\text{dis}, \omega}_{i, t}-p^{\text{ch}, \omega}_{i, t} \eta^{\text{ch}}\eta^{\text{dis}}$.  It's easy to verify that
			$\tilde{p}^{\text{b}, \omega}_{i, t} = p^{\text{b}, \omega}_{i, t}$ and  $\tilde{e}^{\text{b}, \omega}_{i, t+1}=e^{\text{b}, \omega}_{i, t}$ (change the ES operation  at time $t$ but do not affect the subsequent energy state). 
			For the constructed  ES operation, we would have  $\tilde{p}^{\omega}_{i, t} = \tilde{p}^{\text{ch}, \omega}_{i, t} - \tilde{p}^{\text{dis}, \omega}_{i, t} = p^{\text{ch}, \omega}_{i, t} \eta^{\text{ch}} \eta^{\text{dis}} - p^{\text{dis}, \omega}_{i, t} < p^{\omega}_{i, t}$. 
			Similarly,  this implies the constructed solution will yield lower cost and thus contradictions. 
		\end{proof}

\vspace{-8mm}		
\section{Proof of \textbf{Theorem} 1}
		\begin{proof}
			We prove the results by definition.  Suppose there are two disjoint coalitions:  $\mathcal{S}^1, \mathcal{S}^2 \subseteq \mathcal{S}$ and $\mathcal{S}^1 \cap \mathcal{S}^2 = \emptyset$. 
			We assume $\bm{x}^*_{\mathcal{S}^1}, [\bm{y}^{\omega}(\bm{x}^*_{\mathcal{S}^1})]_{\omega \in \Omega} $ and $\bm{x}^*_{\mathcal{S}^2}, [\bm{y}^{\omega}(\bm{x}^*_{\mathcal{S}^2})]_{\omega \in \Omega} $ are the optimal solution of the coalition $\mathcal{S}^1$ and $\mathcal{S}^2$, respectively, 
			Since both the objective function and the constraints of problem \eqref{eq:MainProblem} are linear, it's easy to verify that $(\bm{x}^*_{\mathcal{S}^1}+ \bm{x}^*_{\mathcal{S}^2}, [\bm{y}^{\omega}(\bm{x}^*_{\mathcal{S}^1})+\bm{y}^{\omega}(\bm{x}^*_{\mathcal{S}^2})]_{\omega \in \Omega})$ are viable solution for the coalition $\mathcal{S}^1 \cup \mathcal{S}^2$. Therefore, we have
			\begin{equation}
			\begin{split}
			\nu(\mathcal{S}^1 \cup \mathcal{S}^2) \leq \nu(\mathcal{S}^1 ) + \nu(\mathcal{S}^2)
			\end{split}
			\end{equation}
			
		\end{proof}

\bibliographystyle{ieeetr}
\bibliography{reference}

\begin{thebibliography}{10}

\bibitem{de2016value}
F.~J. De~Sisternes, J.~D. Jenkins, and A.~Botterud, ``The value of energy
  storage in decarbonizing the electricity sector,'' {\em {Applied Energy}},
  vol.~175, pp.~368--379, 2016.

\bibitem{Uitlity-scaleES}
``Utility-scale batteries innovation landscape brief.''
  \url{https://www.irena.org/-/media/Files/IRENA/Agency/Publication/2019/Sep/IRENA_Utility-scale-batteries_2019.pdf}.
\newblock Accessed: 2020-12-24.

\bibitem{yu2020energy}
Y.~Yu, Z.~Cai, and Y.~Huang, ``{Energy Storage Arbitrage in Grid-Connected
  Micro-Grids Under Real-Time Market Price Uncertainty: A Double-Q Learning
  Approach},'' {\em {IEEE Access}}, vol.~8, pp.~54456--54464, 2020.

\bibitem{mallapragada2020long}
D.~S. Mallapragada, N.~A. Sepulveda, and J.~D. Jenkins, ``Long-run system value
  of battery energy storage in future grids with increasing wind and solar
  generation,'' {\em {Applied Energy}}, vol.~275, p.~115390, 2020.

\bibitem{hu2010optimal}
W.~Hu, Z.~Chen, and B.~Bak-Jensen, ``{Optimal operation strategy of battery
  energy storage system to real-time electricity price in Denmark},'' in {\em
  {IEEE PES General Meeting}}, pp.~1--7, IEEE, 2010.

\bibitem{heinrichs2013sharing}
H.~Heinrichs, ``{Sharing economy: a potential new pathway to sustainability},''
  {\em {GAIA-Ecological Perspectives for Science and Society}}, vol.~22, no.~4,
  pp.~228--231, 2013.

\bibitem{lombardi2017sharing}
P.~Lombardi and F.~Schwabe, ``Sharing economy as a new business model for
  energy storage systems,'' {\em {Applied Energy}}, vol.~188, pp.~485--496,
  2017.

\bibitem{koirala2019community}
B.~P. Koirala, R.~A. Hakvoort, E.~C. van Oost, and H.~J. van~der Windt,
  ``{Community energy storage: Governance and business models},'' {\em
  {Consumer, Prosumer, Prosumager: How Service Innovations Will Disrupt the
  Utility Business Model}}, pp.~209--234, 2019.

\bibitem{zhu2019credit}
H.~Zhu and K.~Ouahada, ``{Credit-Based Distributed Real-Time Energy Storage
  Sharing Management},'' {\em {IEEE Access}}, vol.~7, pp.~185821--185838, 2019.

\bibitem{koirala2018community}
B.~P. Koirala, E.~van Oost, and H.~van~der Windt, ``{Community energy storage:
  A responsible innovation towards a sustainable energy system?},'' {\em
  {Applied Energy}}, vol.~231, pp.~570--585, 2018.

\bibitem{fleischhacker2018sharing}
A.~Fleischhacker, H.~Auer, G.~Lettner, and A.~Botterud, ``{Sharing solar PV and
  energy storage in apartment buildings: resource allocation and pricing},''
  {\em {IEEE Transactions on Smart Grid}}, vol.~10, no.~4, pp.~3963--3973,
  2018.

\bibitem{chakraborty2018sharing}
P.~Chakraborty, E.~Baeyens, K.~Poolla, P.~P. Khargonekar, and P.~Varaiya,
  ``Sharing storage in a smart grid: A coalitional game approach,'' {\em {IEEE
  Transactions on Smart Grid}}, vol.~10, no.~4, pp.~4379--4390, 2018.

\bibitem{bayram2015stochastic}
I.~S. Bayram, M.~Abdallah, A.~Tajer, and K.~A. Qaraqe, ``A stochastic sizing
  approach for sharing-based energy storage applications,'' {\em {IEEE
  Transactions on Smart Grid}}, vol.~8, no.~3, pp.~1075--1084, 2015.

\bibitem{kim2017optimal}
I.~Kim and D.~Kim, ``Optimal capacity of shared energy storage and photovoltaic
  system for cooperative residential customers,'' in {\em {2017 International
  Conference on Information and Communications (ICIC)}}, pp.~293--297, IEEE,
  2017.

\bibitem{yao2015optimal}
J.~Yao and P.~Venkitasubramaniam, ``Optimal end user energy storage sharing in
  demand response,'' in {\em {2015 IEEE International Conference on Smart Grid
  Communications (SmartGridComm)}}, pp.~175--180, IEEE, 2015.

\bibitem{yao2017privacy}
J.~Yao and P.~Venkitasubramaniam, ``Privacy aware stochastic games for
  distributed end-user energy storage sharing,'' {\em {IEEE Transactions on
  Signal and Information Processing over Networks}}, vol.~4, no.~1, pp.~82--95,
  2017.

\bibitem{hallefjord1995computing}
{\AA}.~Hallefjord, R.~Helming, and K.~J{\o}rnsten, ``Computing the nucleolus
  when the characteristic function is given implicitly: A constraint generation
  approach,'' {\em {International Journal of Game Theory}}, vol.~24, no.~4,
  pp.~357--372, 1995.

\bibitem{alskaif2017reputation}
T.~AlSkaif, A.~C. Luna, M.~G. Zapata, J.~M. Guerrero, and B.~Bellalta,
  ``Reputation-based joint scheduling of households appliances and storage in a
  microgrid with a shared battery,'' {\em {Energy and Buildings}}, vol.~138,
  pp.~228--239, 2017.

\bibitem{zhang2020service}
W.~Zhang, W.~Wei, L.~Chen, B.~Zheng, and S.~Mei, ``{Service pricing and load
  dispatch of residential shared energy storage unit},'' {\em {Energy}},
  p.~117543, 2020.

\bibitem{zhao2017pricing}
D.~Zhao, H.~Wang, J.~Huang, and X.~Lin, ``Pricing-based energy storage sharing
  and capacity allocation,'' in {\em {IEEE International Conference on
  Communications (ICC)}}, pp.~1--6, IEEE, 2017.

\bibitem{zhao2019virtual}
D.~Zhao, H.~Wang, J.~Huang, and X.~Lin, ``{Virtual Energy Storage Sharing and
  Capacity Allocation},'' {\em {IEEE Transactions on Smart Grid}}, 2019.

\bibitem{liu126optimal}
J.~Liu, X.~Chen, Y.~Xiang, D.~Huo, and J.~Liu, ``Optimal planning and
  investment benefit analysis of shared energy storage for electricity
  retailers,'' {\em {International Journal of Electrical Power \& Energy
  Systems}}, vol.~126, p.~106561.

\bibitem{saad2009coalitional}
W.~Saad, Z.~Han, M.~Debbah, A.~Hjorungnes, and T.~Basar, ``Coalitional game
  theory for communication networks,'' {\em {IEEE Signal Processing Magazine}},
  vol.~26, no.~5, pp.~77--97, 2009.

\bibitem{harsha2014optimal}
P.~Harsha and M.~Dahleh, ``Optimal management and sizing of energy storage
  under dynamic pricing for the efficient integration of renewable energy,''
  {\em {IEEE Transactions on Power Systems}}, vol.~30, no.~3, pp.~1164--1181,
  2014.

\bibitem{pandvzic2018optimal}
H.~Pand{\v{z}}i{\'c}, ``Optimal battery energy storage investment in
  buildings,'' {\em {Energy and Buildings}}, vol.~175, pp.~189--198, 2018.

\bibitem{schulz2013approximating}
A.~S. Schulz and N.~A. Uhan, ``Approximating the least core value and least
  core of cooperative games with supermodular costs,'' {\em {Discrete
  Optimization}}, vol.~10, no.~2, pp.~163--180, 2013.

\bibitem{gillies1959solutions}
D.~B. Gillies, ``Solutions to general non-zero-sum games,'' {\em {Contributions
  to the Theory of Games}}, vol.~4, pp.~47--85, 1959.

\bibitem{schmeidler1969nucleolus}
D.~Schmeidler, ``The nucleolus of a characteristic function game,'' {\em {SIAM
  Journal on applied mathematics}}, vol.~17, no.~6, pp.~1163--1170, 1969.

\bibitem{BuildingDemand}
``Openei.'' \url{https://openei.org/datasets/files/961/pub/}.
\newblock Accessed: 2020-10-29.

\bibitem{WindSolar}
``Measurement and instrumentation data center (midc), nrel transforming
  energy.''
  \url{https://midcdmz.nrel.gov/apps/daily.pl?site=NWTC&start=20010824&yr=2020&mo=1&dy=28}.
\newblock Accessed: 2020-10-29.

\bibitem{shi2018optimal}
Y.~Shi, B.~Xu, Y.~Tan, D.~Kirschen, and B.~Zhang, ``Optimal battery control
  under cycle aging mechanisms in pay for performance settings,'' {\em IEEE
  Transactions on Automatic Control}, vol.~64, no.~6, pp.~2324--2339, 2018.

\end{thebibliography}


%\vfill


\end{document}